\documentclass[A4,amsmath,showpacs,amssymb,aps]{revtex4-1}
\usepackage{hyperref}
\usepackage{graphicx}
\usepackage{dcolumn}
\usepackage{bm}
\usepackage{hyperref}
\usepackage{pdfpages}
\usepackage{pgffor}
\def\Xint#1{\mathchoice
    {\XXint\displaystyle\textstyle{#1}}%
    {\XXint\textstyle\scriptstyle{#1}}%
    {\XXint\scriptstyle\scriptscriptstyle{#1}}%
    {\XXint\scriptscriptstyle\scriptscriptstyle{#1}}%
      \!\int}
\def\XXint#1#2#3{{\setbox0=\hbox{$#1{#2#3}{\int}$}
    \vcenter{\hbox{$#2#3$}}\kern-.5\wd0}}
\def\dashint{\Xint-}
\begin{document}
%--------------------------------------------------------
\newcommand{\beq}{\begin{equation}}
\newcommand{\eeq}{\end{equation}}
\newcommand{\beqar}{\begin{eqnarray}}
\newcommand{\eeqar}{\end{eqnarray}}
\newcommand{\gsim}{\stackrel{>}{\sim}}
\newcommand{\spc}{\mbox{ }}
\newcommand{\dspc}{\mbox{  }}
\pagenumbering{gobble}
%++++++++++++++++++++++++++++++++++++++++++++++++++++++++++++++++++++++++
\title[Compressed Sensing for TF GW Data Analysis  ...]
{Compressed Sensing for Time-Frequency Gravitational Wave Data Analysis}
%++++++++++++++++++++++++++++++++++++++++++++++++++++++++++++++++++++++++
\author{Paolo Addesso, Maurizio Longo, Stefano Marano, Vincenzo Matta}
\affiliation{Dept. of Electrical and Computer Engineering and Applied Mathematics, 
University of Salerno, 84084 Fisciano (SA), Italy}
\author{Maria Principe, Innocenzo M Pinto}
\affiliation{Waves Group, University of Sannio at Benevento, 82100 Benevento, Italy, INFN, LVC and KAGRA}
%--------------------------------------------------------
\date{\today}
%--------------------------------------------------------
\begin{abstract}
The potential of compressed sensing for obtaining sparse  time-frequency representations for gravitational wave data analysis is illustrated by comparison with existing methods, as regards i) shedding light on the fine-structure of  noise transients (glitches) in preparation of their classification, and ii) boosting the performance of  waveform consistency  tests in the  detection of unmodeled transient gravitational wave signals  using a network
of detectors affected by unmodeled noise transients.
\end{abstract}
%--------------------------------------------------------
\maketitle
%
%%%%%%%%%%%%%%%%%%%%%%%%%%%%%%
\section{Introduction and Motivation}
%%%%%%%%%%%%%%%%%%%%%%%%%%%%%%
%
An important  fraction of the Gravitational Wave  (henceforth GW) signals of cosmic origin 
sought after  by running and/or planned GW detection experiments 
based on optical interferometers \cite{LIGO}-\cite{KaGra} 
consists of {\it transient} signals.
These include, e.g., binary inspirals/mergers/ringdowns, and supernova outbursts.\\
Transient  {\it disturbances}  of  instrumental and/or enviromental origin, 
globally nicknamed {\it glitches}, are also ubiquitously observed in the data
of interferometric GW detectors,
occurring with various amplitudes, shapes, and rates.\\
In view of the essentially {\it unmodeled} nature of  transient GWs 
of astrophysical interest \cite{Kokkotas},  
it is challenging to get rid from glitches.
The problem is further complicated by the fact that glitches infringe  
the additive  Gaussian stationary  noise assumption on which
most detection/estimation algorithms are based \cite{Glitch_Bible}.\\ 
Substantial work has been made in  GW experiments
to understand the origin of glitches using additional information from environment
and instrument monitoring channels, and to obtain reliable data quality assessment
and vetoing criteria \cite{DQ1}-\cite{DQN}.  
Consistency tests among the data gathered by 
several non co-located interferometers  have been suggested
and implemented to discriminate glitches, in view of their essentially 
{\it local} nature \cite{network}.\\
In the  time-frequency (henceforth TF) domain, 
where non-stationary  signals are most naturally represented \cite{Cohen},
typical GW transients as well as typical instrumental glitches 
exhibit  highly localized  supports,
and are thus technically  {\it sparse}.
GW chirps from inspiraling binary systems
with large duration $\times$ bandwidth figures
are also sparse in the TF domain, 
yielding almost one-dimensional signatures ({\it ridges}),
representing their instantaneous-frequency evolutions \cite{Boashash}.\\
In the last few years a unifying conceptual framework 
for the efficient representation of  sparse signals 
has been developed, globally referred to under the name of 
{\it compressed sensing}  (henceforth CS).
The power of the CS paradigm is witnessed 
by the exponentially growing scope of its applications 
(see, e.g., \cite{Rice_uni} for a broad and up to date list).\\
Using CS related concepts/tools  in GW data analysis 
we may take advantage of the {\it sparsity} 
of both GW signals and instrumental glitches in the TF domain,
for the purpose of  
i) enhancing the readability of TF representations of the data
gathered by interferometric detectors, and 
ii)  improving consistency tests among different detectors.\\
The potential of sparse representation for GW data analysis
was independently recognized in \cite{our_sparse1}  and \cite{Ra}, 
under a different perspective.\\
Our aim in this paper will be to illustrate  by examples 
the potential of CS  for TF GW data analysis,
extending previous results in \cite{our_sparseN}.\\
The paper is accordingly organized as follows.
In Section \ref{sec:GWandGL} we review 
the key  features of the relevant  signal(s) and noise(s)
and in Section \ref{sec:TF} the basics  of the  TF representations 
referred to in the paper.\\
In Section \ref{sec:WV_to_SK}  we introduce sparse (skeletal) TF representations, 
and propose a CS-based algorithm for constructing them, 
together with pertinent implementation details.\\
In Section \ref{sec:num}  we discuss a number of numerical experiments, 
where the proposed sparse TF representations are evaluated
by comparison with other TF representation tools presently in use in GW data analysis. 
Conclusions follow under Section \ref{sec:concl}.
%
%%%%%%%%%%%%%%%%%%%%%%%%%%%%%%%%
\section{Transient  GW Signals and Glitches}
\label{sec:GWandGL}
%%%%%%%%%%%%%%%%%%%%%%%%%%%%%%%%
%
Transient GW signals can be represented in the form
\beq
s(t)=\mbox{Re}
\left[\tilde{s}(t)\right], \mbox{  where  }
\tilde{s}(t) = S(t) \exp\left[ \imath \psi(t) \right],
\label{eq:x0}
\eeq
where the amplitude $S(t)$  function evolves adiabatically 
(i.e., on much longer timescales) 
compared to the phase $\psi(t)$, so that 
\beq
\left| 
\frac{ \dot{S} }{S} 
\right| 
\ll 
|\dot{\psi}|,
 \mbox{and   } 
\left| \frac{\ddot{\psi}} { \dot{\psi}^2 } \right| \ll 1.
\label{eq:adiabatic}
\eeq
GW signals  emitted  by compact binary systems 
during their inspiral \cite{Inspiral}, 
merger  \cite{Lazarus},
and ringdown \cite{ringdown} phase,
as well as supernova core-collapse GW waveforms \cite{Dimmel}
belong to the above class.\\
The data gathered by interferometric GW detectors are embedded in  additive noise. 
Once all narrowband features (power lines, natural-modes of the test-mass suspending
wires, etc.) have been subtracted,  the residual noise floor is found to consists
of a  locally-stationary, zero mean, coloured Gaussian random process $n(t)$,  
and a  generalized  impulsive  component $g(t)$. 
This latter  is basically  a (random) superposition of transients of environmental/instrumental
origin (known as {\it glitches}), viz. \cite{Principe}: 
\beq
g(t)=\sum_{k=1}^{K} \psi_{k}(t),
\label{eq:glitch_N}
\eeq
$K$ being the (random) number of such transients 
in the observation time-window, and $\psi_k(\cdot)$
the (random) glitches therein. \\ 
Transient GWs as well as glitches are typically  {\it highly  localized} in the TF plane,
their support being restricted to tiny {\it spots} or narrow {\it stripes}. 
% 
%%%%%%%%%%%%%%%%%%%%%%%%%%%%%%%%%%%%%%%%%%%%%%%%%%%
\section{Time-Frequency Representations}
\label{sec:TF}
%%%%%%%%%%%%%%%%%%%%%%%%%%%%%%%%%%%%%%%%%%%%%%%%%%%
%
Time-frequency   representations 
provide the most natural framework  for 
representing  non stationary (transient) signals.
A variety of different TF representations
have been proposed, mostly in the XX century \cite{TFbook_1}-\cite{TFbook_N}.\\
Several have been suggested or used for GW  data analysis,
including the linear short-time-Fourier, 
wavelet and constant-Q transforms
(shortly reviewed in Sections \ref{sec:STFT}-\ref{sec:QT}),
the bilinear Bertrand \cite{Bertrand_TF}
and Wigner-Ville  (section \ref{sec:WV}) transforms,
and the special (sparse) decompositions 
named after Hilbert-Huang (Section \ref{sec:HHT})
and Mallat-Zhang (Section \ref{sec:AD}).\\
More exist, including 
the bilinear hyperbolic \cite{hyper_TF},
higher-order
(e.g., generalized Cohen-class \cite{Cohen_4},
L-Wigner \cite{L_Wigner},
polyspectral \cite{Nikias},
polynomial \cite{poly_WV}) ,
and positive-definite \cite{positive_TFD}
TF representations,
that haven't been used so far in GW data analysis,
to the best of our knowledge.
%
%=============================================
\subsection{Time-Frequency Duality}
\label{sec:GHP}
%=============================================
%
Time-frequency duality (also known as the Gabor-Heisenberg principle)
expresses the well known fact that the wider the time-support of $x(t)$,
the narrower its frequency-support, and vice-versa.
This property, that plays a key role in TF data analysis, 
can be formalized by introducing the time and frequency {\it barycenters} 
\beq
t_0=
\int_{-\infty}^{+\infty}
\!\!
t |x(t)|^2 dt,
\mbox{    }
f_0=
\int_{-\infty}^{+\infty}
\!\!
f |X(f)|^2 df,
\label{eq:centroids}
\eeq
and the related {\it spreads} $\delta^{(x)}_t$ and $\delta^{(x)}_f$,
\beq
\delta^{(x)}_t=
\left[
\int_{-\infty}^{+\infty}
\!\!
(t-t_0)^2 |x(t)|^2 dt 
\right]^{1/2},
\mbox{    }
\delta^{(x)}_f=
\left[
\int_{-\infty}^{+\infty}
\!\!
(f-f_0)^2 |X(f)|^2 df 
\right]^{1/2},
\label{eq:spreads}
\eeq
$x(t)$ and $X(f)$  being a generic time function and its Fourier transform,  
yielding \cite{Cohen}
\beq
\delta^{(x)}_t \delta^{(x)}_f = \xi_x \geq (4\pi)^{-1}.
\label{eq:GaborHeisenberg}
\eeq
The quantity  $\xi_x$ depends on the chosen $x$ function,
is known as its TB  (time-bandwidth) product,
and attains its lower bound when $x(t)$ is a (unit-norm) Gabor function
\beq
x(t)=
\frac{1}
{\sqrt[4]{2\pi [\delta^{(x)}_t]^2}}
\exp\left(-2\pi\imath f_0 t \right)
\exp
\left[-
\left(
\frac{t-t_0}{2 \delta^{(x)}_t}
\right)^2
\right].
\label{eq:GaborFun}
\eeq 
%
%=============================================
\subsection{Short Time Fourier Transform}
\label{sec:STFT}
%=============================================
%
\noindent
The short-time Fourier transform (STFT) is likely the simplest and oldest
TF representation, and is defined by:
\beq
{\cal F}^{(h)}_x (t,f)=
\int_{-\infty}^{+\infty}
\!\!
x(\tau)h(t-\tau)\exp(-2\pi\imath f \tau) d\tau
\label{eq:STFT}
\eeq
where $h(t)$ is a time-windowing function which vanishes 
for $|t|>\delta^{(h)}_t$. 
The STFT is the projection of $x(t)$ into an analyzing function 
whose envelope is a time-shifted version of $h(\tau)$  centered at $\tau=t$, 
whose (complex) carrier has frequency $f$.   
The (squared) modulus of the STFT, known as {\it spectrogram}, 
\beq
S^{(h)}_x (t,f)=
\left|
{\cal F}^{(h)}_x (t,f)
\right|^2
\label{eq:SPG}
\eeq
has been frequently used as a fiducial representation of the 
frequency-distribution of signal energy for non-stationary signals.
The frequency resolution of the STFT (and spectrogram) is the same for {\it all}  $f$.
Time-frequency duality implies that smaller values of $\delta^{(h)}_t $  
(yielding larger time localization) entail poorer frequency resolution. 
%
%=============================================
\subsection{Wavelet Transform}
\label{sec:WLT}
%=============================================
%
\noindent
The wavelet transform (WT) is defined by:
\beq
{\cal D}^{(h)}_x (t,f)=
\left(\frac{f}{f_h}\right)^{1/2}
\int_{-\infty}^{+\infty}
\!\!
x(\tau)
h^{*}\left[\frac{f}{f_h}\left(t-\tau \right)\right] d\tau.
\label{eq:WT}
\eeq
Here $h(t)$  is  a  zero-mean complex-valued  time-windowed oscillatory function
whose absolute Fourier spectrum is unimodal and peaked at $f=f_h$, 
known as  {\it mother wavelet}, viz.,
\beq
h(t)=w(t) \exp\left(-2\pi\imath f_h t \right), 
\mbox{  where    } 
w(t) = 0, \mbox{  } \forall |t|>\delta^{(w)}_t.
\label{eq:motherWav}
\eeq
It is seen that the wavelet transform is the projection of $x(\tau)$ 
into an analyzing function obtained by time-shifting
the mother wavelet (from $\tau=0$ to $\tau=t$),
and {\it time-scaling} it via the frequency-dependent factor  $a=(f/f_h)$
(which entails frequency-shifting its carrier from $f_h$ to $f$). \\
Depending on whether $a \lessgtr 1$, the above time-scaling 
corresponds to {\it time-squeezing}  or {\it time-stretching},
and at the same time, in view of  TF duality, 
to {\it frequency-stretching} or {\it frequency-sequeezing},
respectively, by the same factor $a$.\\
The (squared) modulus of the WT, known as {\it scalogram},
\beq
\Sigma^{(h)}_x (t,f)=
\left|
{\cal D}^{(h)}_x (t,f)
\right|^2
\label{eq:SCG}
\eeq
can be used as a fiducial representation of the TF distribution of signal energy.
We note in passing that both the spectrogram (\ref{eq:SPG}) and the scalogram (\ref{eq:SCG}) 
are unitary representations, i.e., satisfy signal energy conservation (Parseval theorem), viz.,
\beq
\int_{-\infty}^{+\infty}
|x(\tau)|^2
d\tau=
\int_{-\infty}^{+\infty}
\!\!\! dt 
\int_{-\infty}^{+\infty}
\!\!\! df
\mbox{ }S^{(h)}_x (t,f)
=
\int_{-\infty}^{+\infty}
\!\!\! dt
\int_{-\infty}^{+\infty}
\!\!\! df
\mbox{ }\Sigma^{(h)}_x (t,f).
\eeq
%
%=============================================
\subsection{Q-Transform}
\label{sec:QT}
%=============================================
%
\noindent
As noted above, the STFT (and the spectrogram) 
is a {\it constant-bandwidth} representation, 
where the analyzing function duration (and bandwidth) 
is {\it the same} for all $(t,f)$.
In the WT (and the scalogram), on the other hand, 
the effective time-width  $\delta_t$
of the analyzing function at $f$
is inversely proportional to  $f$, 
and hence (in view of Gabor-Heisenberg theorem) 
the spectral-width $\delta_f$ is proportional to $f$, 
so that the ratio $f/\delta_f$
is {\it the same}  for all $f$.  
Hence, the WT (and the scalogram) 
are basically  {\it constant-Q}  representations,
$Q$ being the name given to  the $f/\delta_f$ ratio. \\
A different, but closely related, constant-Q  TF representation 
is obtained  by  starting from the STFT , eq. (\ref{eq:STFT}),
and letting the time-width of the windowing function
to be inversely proportional to $f$, viz.:
\beq
\delta_t = \frac{C}{f}
\eeq
so that, in view of TF duality
\beq
\delta_f=\frac{\xi}{\delta_t}=\frac{f \xi  }{C}
\eeq
yielding
\beq
\frac{f}{\delta_f}=:Q=\frac{C}{\xi}.
\eeq
Under this assumption, equation (\ref{eq:STFT}) yields  
the so called {\it constant-Q transform} (henceforth QT), 
\beq
{\cal Q}^{(h)}_x(t,f;Q) = 
\int_{-\infty}^{\infty}
x(\tau)
h_Q(t-\tau;f)
\exp(-\imath 2\pi f \tau)
d\tau 
\label{eq:QT1}
\eeq
where,  denoting as  $h(t)$ a chosen windowing function 
with time width  $\delta^{(h)}_t$  and  TB figure  $\xi_h$,  
\beq
h_Q(t-\tau;f) = h\left[ \frac{f \delta^{(h)}_t}{Q \xi_h} (t-\tau) \right].
\eeq
The form of eq.  (\ref{eq:QT1}) suggests a convenient  way 
for its computation, via a sequence of Fourier transforms
(denoted below by the ${\cal F}$ operator), viz.:
\beq
{\cal Q}^{(h)}_x(t,f;Q)=
{\cal F}^{-1}_{\xi \rightarrow t}
\left[
X(\xi,f)H_Q^{*}(\xi;f)
\right]
\label{eq:QT1a}
\eeq
where,
\beq
X(\xi,f)={\cal F}_{t \rightarrow \xi}
[x(t)\exp(-2\imath\pi f t)],
\eeq
and
\beq
H_Q(\xi,f)={\cal F}_{t \rightarrow \xi}
[h_Q(t;f)].
\label{eq:QT2}
\eeq
The QT  tiles  the TF plane  linearly in time  and logarithmically in frequency,
as shown in Figure 1, by comparison with the STFT and WT.
In exploratory TF data analysis, 
the QT is  usually computed for a set 
of  (logarithmically-spaced) Q values,
lower (higher) Qs yielding lower (higher) resolution in frequency, 
and higher (lower) resolution in time. 
The  QT  was introduced and developed in \cite{Brown_1}, \cite{Brown_2},
and was proposed as a tool for GW data analysis in \cite{QT0}, \cite{QT1},
leading to several currently perused implementations in LIGO-Virgo, 
including the the {\it Q}, {\it Omega}, and {\it Omicron} 
pipelines \cite{QT0}-\cite{Omicron}.
%
%
%%%%%%%%%%%%%%%%%%%%%%%%%%%%%%%%%%%%%%%%%%%%%%%%%%%
\subsection{Wigner-Ville Transform}
\label{sec:WV}
%%%%%%%%%%%%%%%%%%%%%%%%%%%%%%%%%%%%%%%%%%%%%%%%%%%
% 
The Wigner-Ville (henceforth WV) transform  of a real valued signal  $x(t)$  reads \cite{Claasen}:
\beq
W_x(t,f)=\int_{-\infty}^{+\infty} 
\tilde{x}\left(t+\frac{\tau}{2}\right)
\tilde{x}^\ast\left(t-\frac{\tau}{2}\right)
e^{-i2\pi f \tau}d\tau
\eeq
where  $\tilde{x}(t)$   is the the analytic mate  of  $x(t)$,
\beq
\tilde{x}(t)=x(t)+i\,{\cal H}[x(t)],
\eeq
${\cal H}[\cdot]$ denoting the Hilbert transform operator,
\beq
{\cal H}[x(t)]
=
\frac{1}{\pi}
\dashint_{R}
\frac{x(\tau)}{t-\tau} d\tau 
=
{\cal F}^{-1}
\left\{
U(f){\cal F}[x]
\right\}
\eeq
where $U(\cdot)$ is Heaviside's step function.\\ 
Using the WV has been suggested for different purposes in GW  data analysis, 
including the detection of GW chirps from inspiraling binaries,
\cite{Feo}-\cite{Pai},
and the estimation of GW arrival time-delays in a network of detectors
for source localization \cite{XWV}.\\ 
The WV has a number of nice (and unique) properties \cite{Cohen}
among all time-frequency representations:
it is a member of the Cohen Class (i.e., it is covariant 
with respect to time and/or frequency shifts of its argument); 
its time (frequency) barycenter at fixed frequency (time)
reproduce (and define) the group-delay and instantaneous frequency of its argument; 
its marginal distribution along any radial line in the TF plane yields
the energy density of the corresponding fractional Fourier transform \cite{marginals}, 
\cite{footnote3}.
The WV is also {\it unitary}, i.e. energy preserving (Moyal theorem),
\beq
\left[
\int_{R} |\tilde{x}(t)|^2 dt
\right]^2 =
\iint_{R^2} |W_x(t,f)|^2 dt df
\eeq
and invertible (up to an irrelevant complex factor):
\beq
\tilde{x}(t)=\frac{1}{\tilde{x}^*(0)}
\int_{-\infty}^{\infty}
W_x(t/2,f)\exp(2\imath\pi f t) df.
\eeq
The main limitation  of the WV  stems from it bilinear nature, entailing 
in general the appearance of intermodulation artifacts, which
hinder its visual readability.
The WV transform is  immune from such artifacts 
only  in  two special cases, 
namely,  when  $x(t)$ is  either a single noise-free constant-amplitude chirp 
whose frequency changes {\it linearly} with time,  
or a single noise-free  Gabor (sine-Gaussian) function
\cite{Claasen}.\\
Different strategies  have been proposed 
to get rid of the WV intermodulation artifacts,
such as  smoothing kernels \cite{Hlawatsch}
%and ambiguity function weighting~\cite{Flandrin2}, 
and  reassignment~\cite{FlandrinMottin},
that are briefly discussed below. \\
%
%+++++++++++++++++++++++++++++++++++
\subsubsection{Smoothed WV}
%+++++++++++++++++++++++++++++++++++
%
The 2D Fourier transform of the Wigner-Ville distribution
\beq
A_x(\xi,\eta)={\cal F}_
{\tiny {\begin{array}{l}
t \! \rightarrow \! \xi \\ f \! \rightarrow \! \eta 
\end{array}}}
[W_x(t,f)]
\label{eq:AF}
\eeq
%where ${\cal F}$ is the 2D-Fourier  operator, 
is known as  the  Ambiguity (or Woodward) Function (henceforth AF) of $x$,
and the  $(\xi,\eta)$ plane is referred to as the AF domain,
the $\xi$ and $\eta$ arguments being referred to as 
the Doppler-shift  $[\mbox{sec}^{-1}]$ and the delay $[\mbox{sec}]$. \\
Interference artifacts in the WV exhibit rapid variations in the TF plane. 
Accordingly, they map far away from the origin in the 
AF plane, and can be effectively suppressed 
by multiplying  the  AF  by some low-pass 2D-window function $\Lambda(\xi,\eta)$ 
vanishing beyond some distance from the origin of the AF plane,
and transforming the windowed AF back to the  $(t,f)$  domain. \\
Multiplication in the $(\xi,\eta)$ domain corresponds 
to convolution in the $(t,f)$ domain (Borel theorem),
and the resulting smoothed WVD can be  written:
\beq
W^{(\lambda)}_x(t,f)=
\iint_{R^2}
W_x(\tau,\eta)\lambda(t-\tau,f-\nu)d\tau d\nu
\label{eq:smoothedWV}
\eeq
where $\lambda(t,f)$  is the inverse Fourier transform
of the 2D-windowing factor $\Lambda(\xi,\eta)$ 
in the AF domain, and is referred to as {\it smoothing kernel}.
Smoothing (low pass filtering) of the WV entails some loss in its TF resolution.\\ 
It is worth noting that both the spectrogram (\ref{eq:SPG}) and the
scalogram (\ref{eq:SCG}) are special smoothed versions of the WV,
where the smoothing  kernel  $\lambda(t,f)$  
is the WV transform of the pertinent windowing functions $h$  
in eqs. (\ref{eq:STFT}) and (\ref{eq:WT})
\cite{Hlawatsch}.\\
Several smoothing kernels have been proposed, featuring different
properties \cite{HlawatschSmooth}r.
In a series of papers \cite{BJ1}-\cite{BJN}  Baraniuk and Jones, 
developed the idea of seeking a radially-Gaussian 
windowing function in the AF plane, often referred to as radially
Gaussian kernel (RGK) in the technical Literature,
tailored to the actual energy distribution in the AF plane.
They accordingly use polar coordinates $(r,\theta)$ in the AF plane, and let  
\beq
K(r,\theta)=\exp\left[-\frac{r^2}{2\sigma^2(\theta)}\right]
\label{eq:radial}
\eeq
where, in view of the  AF symmetry property $A_x(-\xi,-\tau)=A_x(\xi,\tau)$,
\beq
K(r,\theta+\pi)=K(r,\theta),
\label{eq:symm_const}
\eeq
and  seek  $\sigma(\theta)$ 
so as to maximize the energy content of the kernel-weighted AF
\beq
\int_{0}^{2\pi} d\theta
\int_{0}^{\infty} r dr
\left|
A_x(r,\theta)K(r,\theta)
\right|^2
\label{en_cont}
\eeq
subject to a measure (area) constraint:
\beq
\int_{0}^{2\pi} d\theta
\int_{0}^{\infty} r dr
\left|
K(r,\theta)
\right|^2
\leq 
\alpha.
\label{eq:vol_const}
\eeq
The Baraniuk-Jones radially Gaussian kernels 
are quite effective 
in producing intermodulation-artifact free smoothed versions of the WV, 
with nicely limited (and uniform) loss in TF resolution.
The RGK smoothed version of the WV will be referred to 
as BJ-smoothed WV in the rest of this paper.
%
%++++++++++++++++++++++++++++++++++++++++
\subsubsection{Reassignment} 
\label{sec:reassign}
%++++++++++++++++++++++++++++++++++++++++
%
Reassignment is, basically, a heuristic procedure for {\it re-focusing}
the WV, after it was blurred  by smoothing,
and can be understood without referring to a particular
smoothing kernel \cite{FlandrinMottin}.\\
According to (\ref{eq:smoothedWV}), the value of the smoothed WV 
at any given point $(t,f)$ is a weighted sum 
of all $W_x$  values throughout the TF plane,  
the weighting factor being represented by the 
function $\lambda(\cdot,\cdot)$ 
centered in $(t,f)$.\\
Following Flandrin  (who re-discovered and elaborated
the reassignment concept introduced in \cite{proto_reassign}), 
''it is as if the total mass of an object were assigned 
to its geometric center - which is incorrect, except in the 
special case of {\it homogeneous} density''.
Reassignment accordingly consists in computing the 
TF coordinates of the {\it centroid} of $W_x$, 
as weighted by the $(t,f)$-centered $\lambda(\cdot,\cdot)$,
viz.:
\beq
\hat{t}^{(h)}_x=\frac{
\iint_{R^2}
\tau W_x(\tau,\eta)\lambda(t-\tau,f-\nu)d\tau d\nu
}{
\iint_{R^2}
W_x(\tau,\eta)\lambda(t-\tau,f-\nu)d\tau d\nu,
}
\eeq
\beq
\hat{f}^{(h)}_x=\frac{
\iint_{R^2}
\nu W_x(\tau,\eta)\lambda(t-\tau,f-\nu)d\tau d\nu
}{
\iint_{R^2}
W_x(\tau,\eta)\lambda(t-\tau,f-\nu)d\tau d\nu
}
\eeq
and {\it reassigning} the value of the WVD
originally computed in $(t,f)$ 
to the point $(\hat{t}^{(h)}_x, \hat{f}^{(h)}_x)$.
Reassignment performs very well in the absence of noise. 
%
%============================================ 
\subsection{Hilbert-Huang-Transform}
\label{sec:HHT}
%============================================ 
%
\noindent
The Hilbert-Huang-Transform (henceforth HHT), consists of two steps. 
The first one, known as Empirical Mode Decomposition (EMD) 
\cite{Huang_EMD}, 
is a constructive recipe for adaptively representing a
nonstationary signal as 
\beq
S(t)=\sum_{k=1}^N s_k(t) + r(t)
\label{eq:EMD}
\eeq
where $r(t)$ is a {\it monotonic} (possibly null) residual,
and the {\it intrinsic mode functions} (IMF) $s_k(t)$ 
are obtained by an exhaustive {\it sifting} procedure, 
consisting in defining the functions 
\beq
m_{h-1}(t)=\left\{
\begin{array}{l}
S(t),\mbox{  }h=1,\\
\\
\displaystyle{
S(t)-\sum_{k=1}^{h-1} s_k(t)
},\mbox{  }h>1,
\end{array}
\right.
\label{eq:sift1}
\eeq
computing their {\it mean-envelope} $\mu_{h-1}$ 
(average of upper and lower envelopes),
and letting  
\beq
s_{h}=m_{h-1}-\mu_{h-1}.
\label{eq:sift2}
\eeq
As shown in \cite{EMD_wavelet}, EMD attempts (and often succeeds) 
identifying the various scales at which a signal oscillates, 
in a fully data-driven way.
Indeed, the IMF spectra organize spontaneously as an almost constant-Q 
dyadic wavelet-like filter bank \cite{EMD_constant_Q_bank}.
\\
In the second step, the analytic mate of {\it each} IMF is constructed
via the Hilbert transform (HT), 
\beq
\tilde{s}_k(t)=A_k(t)\exp[\imath\psi_k(t)]=s_k(t)+\imath{\cal H}[s_k(t)]
\eeq
whereby an instantaneous amplitude $A_k(t)$ (henceforth IA) 
is associated to a fiducial {\it instantaneous angular frequency} $\dot{\psi}_k(t)$
(henceforth IF), computed from the (numerical) time-derivative 
of the instantaneous phase $\psi_k(t)$.
This yields the  Hilbert spectrum, aka the HHT:
\beq
H_S(t,\omega)=\frac{1}{2\pi}\sum_k A_k(t) 
\delta\left[f-\frac{\dot{\psi}_k(t)}{2\pi }\right]
\label{eq:Hs}
\eeq
whereby each IMF is represented by a $1D$ feature in the TF plane
(its fiducial instantaneous-frequency line, henceforth IFL), 
whose points have different levels, given by the pertinent IA.\\
The HHT was introduced in GW data analysis in \cite{HHT_GW1}, and 
further exploited in \cite{HHT_GW2}. A recent review 
of the related implementation aspects 
can be found in \cite{HHT_GW3}.\\
Note that the Hilbert spectrum (\ref{eq:Hs}) holds no information 
about the instantaneous bandwidth  (IBW) of the signal 
\cite{IBW_Cohen_1}-\cite{Loughlin_IBW}.
An IBW-aware version of the HHT can be obtained in principle
by estimating the IBW at each point (time) along the IFL, 
and re-distributing the local energy  (squared IA) 
over an IBW-wide frequency interval, 
according to some suitable/fiducial fall-off law.\\
The original EMD method 
lacks a rigorous mathematical foundation, 
and hence its convergence properties, and resilience against noise 
are not well established \cite{Huang_EMD}.\\
In principle, it should be  noted that  
any given (smooth, limited) signal $x(t)$ can be represented 
by an {\it infinite} number of possible $\{a(t),\phi(t)\}$  pairs 
such that $x(t)=a(t)\cos[\phi(t)]$ 
(see, e.g., \cite{Picinbono} for a general discussion, 
and \cite{Boash_ex} for simple examples).
In addition, the HHT algorithm is based on the assumption that
${\cal H}[A(t)\cos[\psi(t)]] = A(t)\sin[\psi(t)]$,  
which holds only if i) the Fourier spectra of the envelope $A(t)$ 
and carrier $\cos[\psi(t)]$ do not nonoverlap \cite{Bedrosian}, 
and ii) $H[\cos[\psi(t)]] = \sin[\psi(t)]$, 
which is (asymptotically) true under broad 
but {\it not} completely general assumptions \cite{Nuttall}.\\
A mathematically sound (and better performing, 
e.g., in the case of multi-tone signals) 
EMD-like algorithm, known as the synchro-squeezed wavelet transform 
was introduced in \cite{Daubechies_synchrosqueezing}.\\
Conceptual as well as technical (implementation) issues  also exist 
as regards both the EMD and the HT steps of the HHT algorithm:
\begin{itemize}
\item{In the original EMD algorithm \cite{Huang_EMD}  
spline-fitting is used to construct  the upper/lower waveform envelopes, 
at each step of the sifting procedure.
This choice entails the appearance of end-point over/undershoot, 
that may hinder faithful IMF recovery. 
Alternative EMD algorithms (based on constrained optimization, 
rather than spline-fitting) have been proposed to circumvent this problem \cite{OPT_IMF}.
Mode mixing, whereby a single IMF may include different signal components, 
or a single component may be split across several IMFs  is another known issue.
Modified EMD algorithm have been used to mitigate this problem,  
the key idea being that of averaging different IMFs obtained by applying EMD 
to several superpositions of the signal with different (independent) realizations of noise \cite{noise_assisted_EMD}. 
Better IMF reconstruction has been achieved using wavelet-based projections 
using the Fej´{e}r-Korovkin class of wavelet filters 
\cite{MODWT_IMF_search}.}
\item{The Hilbert Transform step is usually implemented 
via (fast) discrete Fourier transform (DFT), 
and is accordingly affected by spectral leakage and distorsion, 
which may spoil the subsequent IF estimation.
Numerical differentiation used to retrieve the instantaneous frequency 
is, in addition, quite sensitive to noise.
Alternatives to the DFT-HT have been suggested to fix these limitations, 
with partial success (e.g., interpolative approaches to retrieve  the IAs \cite{LMI_HT}, 
and 1st-order autoregressive modeling for estimating the IFs \cite{AR_alternative_to_HT}). }
\end{itemize}
None of the above HHT improvements has been applied, 
to the best of our knowledge, in GW data analysis so far.\\
The most serious limitation of the HHT lies in its {\it limited} frequency-resolution
(in contrast to claims  made in \cite{HHT_GW1}-\cite{HHT_GW3}). 
Indeed, as shown in  \cite{Flan_EMD1}, a signal consisting of two (pure) tones,  
will {\it not} be solved by the HHT, if the frequency difference
does not exceed a {\it confusion bandwidth} depending on the ratio 
between the amplitudes of the two tones \cite{one_tone}
An improved EMD algorithm with tunable frequency resolution
was proposed in \cite{var_res_EMD}. \\
In \cite{BoaEMD} the  EMD was used,
to derive a set of ambiguity-domain smoothing filters (one for each IMF)
whereby different "views" of the WV of the original signal 
could be obtained, and combined to give a TF representation
referred to as EMD-smoothed WV.\\
%
%%%%%%%%%%%%%%%%%%%%%%%%%%%%%%%%
\subsection{Atomic Decompositions}
\label{sec:AD}
%%%%%%%%%%%%%%%%%%%%%%%%%%%%%%%%
%
Mallat and Zhang are are credited for inventing the matching-pursuit algorithm \cite{Match_Purs}
for decomposing a signal into elementary transients, represented by Gabor atoms 
\cite{Gabor}, \cite{Bastiaans}. \\
The analytic mate of a SG (Gabor) atom $s(t)$  can be approximated by
\beq
\tilde{s}(t)=a(t)\exp[\imath\phi(t)],
\eeq
with
\beq
a(t)=A_0 \exp\{-[(t-t_0)/\Delta T]^2\}, \mbox{       } \phi(t)=2\pi f_0 t+\psi,
\label{eq:SG}
\eeq
provided  the product between the  SG {\it carrier} frequency
$f_0$ and time-spread $\Delta T$ is large ($\gg 1$).
The BT figure of Gabor atoms  attains  the minimum value allowed by
time-frequency duality for any values of the atom parameters, and 
the WV transform of  the SG-Gabor atom (\ref{eq:SG})  is real and positive,
\beq
W_{\tilde{s}}(t,f)\!=\!\Delta T\!\sqrt{2 \pi} A_0^2
\exp \{-2[(t\!-\!t_0)/\Delta T]^2\}
\exp \{-2 \pi^2 [(f\!-\!f_0) \Delta T]^2\}.
\label{eq:WVG}
\eeq
Adding the WVs of the {\it individual} atoms in the AD
yields  a sparse,  intermodulation artifact-free, 
fiducial energy-distribution of the signal in the TF domain, 
known as {\it W(i)V(i)gram} \cite{Match_Purs}.\\
The AD concept was introduced in GW data analysis in \cite{Principe} 
with reference to glitches,
and further pursued in \cite{Cornish_I}, \cite{Cornish_N}.
Under the name of Sine-Gaussians  (henceforth SG),  Gabor atoms
have been also used  to represent generic GW bursts  \cite{SG_GW}.\\
It should be noted that ADs are  {\it non}-unique, 
the choice of the atoms being largely arbitrary.
Alternative atom choices  include, e.g.,  
exponentially damped sinusoids \cite{Prony_like},
and  modulated Gamma-envelopes \cite{Gamma_env}.\\
The physical readability of atomic representations 
is strongly affected by the choice of the atoms, 
and this ultimately leads to the problem of finding the atoms 
which are the most {\it natural} choice for the signals being analyzed 
\cite{which_atoms}.\\
MP-derived ADs may display {\it strong} atoms 
in time intervals where the original signal is {\it negligibly small}. 
These atoms usually {\it cancel out} in the 
reconstructed waveform by destructive interference, 
but stand out in the WiVigram (due to the positive sign
definiteness of \ref{eq:WVG}). 
Pictorially, such atoms are said to form the {\it dark energy} 
of the AD \cite{dark_energy}.\\
Modified MP algorithm have been proposed attempting  to minimize not only 
the reconstruction energy ($L^2$) error,  but also its dark-energy content
\cite{min_dark_MP}, \cite{min_dark_MP2}. 
% 
%%%%%%%%%%%%%%%%%%%%%%%%%%%%%%%%%%
\section{From WV to TF Skeletons via CS}
\label{sec:WV_to_SK}
%%%%%%%%%%%%%%%%%%%%%%%%%%%%%%%%%%%
%
The TF representation of the data  gathered by an interferometric GW detector
contains (noise-blurred)  {\it sparse} features.
These include highly localized {\it juts}, representing  glitches and GW  transients 
with small time-bandwidth figures,
and {\it ridges},  i.e.  almost one-dimensional  features representing 
the frequency evolution of wideband GW chirps.
Together, the above sparse features 
form the  (generalized)  TF {\it skeleton} of  the data.\\
Unfortunately, {\it no} TF representation 
(including the WV,  wavelet and Q transform) 
returns the skeleton of the data.\\
The AD and HHT attempt to find sparse TF approximations consisting 
only of  "juts" or "ridges" , respectively. 
This is physically  unjustified, and limits their range of meaningful applicability.\\
In this section we use the compressed sensing paradigm to derive
general, {\it sparse} and {\it highly resolved} TF representations
from the WV without making any unjustified assumption,
following the approach originally proposed in \cite{FlanBorgn}.\\
As shown in \cite{FlanBorgn}, this technique compares favorably 
to \cite{Hlawatsch}-\cite{FlandrinMottin}
in terms of effectiveness and computational burden.
It is worth stressing, however,  that  it does not  attempt  to  derive 
an artifact-free WV representation, but rather to  {\it construct the skeleton} of the
TF distribution which is a-priori unknown, and otherwise unavailable.\\
The CS paradigm relies on the fact that 
signals that are sparse in  the TF  domain  can be essentially recovered 
using a relatively {\it small} set of samples  from the Fourier-{\it conjugate} domain, 
where the signal is expected to be {\em dense}, according to the Gabor-Heisenberg 
principle \cite{Donoho}, \cite{Candes}.\\
To exploit this property, we start from the AF :
\beq
A_x(\xi,\tau)={\cal F}[W_x(t,f)]
\label{eq:AF1}
\eeq
and manage to distill the (sparsest) TF representation (skeleton) of our data 
using a bunch of values of the AF 
from a suitable neighbourhood  $\Omega$ of  the  origin
in the $(\xi, \tau)$ plane.
Doing so we use again the well known  property of the WV representation
whereby  its highly-oscillatory intermodulation artifacts  
map {\it far away} from the origin in the AF plane .\\
Formally, we seek a {\it sparse} TF distribution $\widehat{\cal W}$ such that
\beq
\widehat{\cal W}(t,f)\!=\!\arg\min||{\cal W}(t,f)||_0:
S_{\Omega}(\xi,\tau)
\left\{
{\cal F}[{\cal W}(t,f)]\!-\!A_x(\xi,\tau) 
\right\}\!=\! 0.
\label{eq:L0}
\eeq
where the $L_0$ norm  $||\cdot||_{0}$  
is the cardinality of the (discrete) support of its argument, and 
\beq
S_{\Omega}(\xi,\tau)=
\left\{
\begin{array}{l}
1,\mbox{ }(\xi,\tau) \in \Omega\\
0,\mbox{ }(\xi,\tau) \notin \Omega
\end{array}.
\right.
\eeq
The {\it cardinality}  (size)  and  {\it shape} of $\Omega$  
should be judiciously chosen for best performance,
as  further discussed in Section \ref{sec:omega}.
%
%%%%%%%%%%%%%%%%%%%%%%%%%%%%%%%%%%%%%%%%%%%%%%%%%%%
\subsection{Computational Cost and  Practical Implementation}
\label{sec:imple}
%%%%%%%%%%%%%%%%%%%%%%%%%%%%%%%%%%%%%%%%%%%%%%%%%%%
%
The optimization problem (\ref{eq:L0}) is combinatorially complex, 
and thus almost unaffordable from a computational viewpoint.
Remarkably, under fairly general conditions
one can rather address the {\it viable} problem
\beq
\widehat{\cal W}(t,f)\!=\!\arg\min||{\cal W}(t,f)||_1:
S_{\Omega}(\xi,\tau)
\left\{
{\cal F}[{\cal W}(t,f)]\!-\!A_x(\xi,\tau) 
\right\}\!=\! 0
\label{eq:opt1}
\eeq
where the $L_1$ norm has replaced the $L_0$ one  \cite{Donoho} .
In order to take into account the noisyness of the data  it is further expedient 
to relax the equality constraint in (\ref{eq:opt1}), replacing it with
a suitable bound on the error  $L_2$ norm, viz.
\beq
\widehat{\cal W}(t,f)\!=\!\arg\!\min||{\cal W}(t,f)||_1\!:\!
|| 
S_{\Omega}(\xi,\tau)\!
\left\{\!
{\cal F}[{\cal W}(t,f)]\!-\!A_x(\xi,\tau) \! 
\right\}\!
||_2 \! \leq \! \epsilon.\mbox{     } 
\label{eq:opt2}
\eeq
The optimization problem  (\ref{eq:opt2}) can be implemented in several ways,
including, e.g., log-barrier, interior-point and iteratively-reweighted least-squares algorithms, whose implementations and computational costs  have been discussed in \cite{L1Magic}, \cite{BP}, \cite{StableRec}, respectively.\\
Convex analysis also tells us that any nontrivial solution of (\ref{eq:opt2}) is  also a solution of
\beq
\widehat{\cal W}(t,f)\!=\!\arg\!\min \lambda 
||{\cal W}(t,f)||_1 \!+\! \frac{1}{2} \!
||\! 
S_{\Omega}(\xi,\tau)\!
\left\{\!
{\cal F}[{\cal W}(t,f)]\!-\!A_x(\xi,\tau) \! 
\right\}\!
||_2^2 \mbox{     }
\label{eq:opt4}
\eeq
for {\it some} $\lambda > 0$, as shown, e.g., in \cite{Fraccanzano}. 
In \cite{Figueiredo} a class of numerically efficient algorithms is proposed
for solving (\ref{eq:opt4}), which are computationally cheaper than solving (\ref{eq:opt2}) directly.
As noted in \cite{Figueiredo}, a sensible choice for $\lambda$  is
\beq
\lambda=\gamma \displaystyle{||{\cal F}^{-1} [\Omega  A_x] ||}_{\infty}, \mbox{  } 
\gamma \sim 0.1
\label{eq:guess}
\eeq 
typically yielding $\epsilon \sim 10^{-2 }$ in (\ref{eq:opt2}) in our numerical experiments,
which is adequate for our present purposes.
%
%%%%%%%%%%%%%%%%%%%%%%%%%%%%%%%%%%%%
\subsection{Choice of the $\Omega$ Domain}
\label{sec:omega}
%%%%%%%%%%%%%%%%%%%%%%%%%%%%%%%%%%%%
%
The choice of the {\it cardinality}  (number of TF samples) and {\it shape} of $\Omega$,  
affects  the skeleton appearance  in rather subtle ways, 
depending on the features of the signals.\\
As the cardinality (hence the diameter) of   $\Omega$  increases, we may expect 
more and more intermodulation products to appear in ${\cal W}$,  
until eventually the whole WV distribution is recovered  
in the limit where $\Omega$ covers the whole  TF plane.
On the other hand, as the cardinality of $\Omega$ is decreased, we expect 
the skeletal components to be gradually spoiled.\\
In \cite{FlanBorgn} a number of multicomponent signals with {\it known} skeletons was considered,
and it was shown that for a WV of size $N \times N$, the choice
$\mbox{card}[\Omega] \sim N$   (the Heisenberg-Gabor cardinality)
was the best one \cite{footnote0}.
However, as noted in \cite{FlanBorgn},  for a given cardinality $N$,
a better skeleton reconstruction  is obtained if the {\it shape} of $\Omega$
{\it fits} the shape of the ambiguity function,
reflecting the  {\it coherency} properties of the signal. \\ 
In this connection it would be desirable to find a systematic  and,
as far as possible,  {\it non}-parametric procedure,
to  determine  the shape of $\Omega$.\\
We adopted a simple (and rather natural) choice for $\Omega$, namely  
the domain whose boundary  $\partial\Omega$  is the  contour level 
of  the RGK  (\ref{eq:radial}) embracing  $N$ samples of the AF.
As we shall see, this choice  yields a more faithful representation 
of  the skeleton, in all cases where this latter is known a priori,
compared to the {\it isotropic} domain  \cite{footnote2}
suggested in \cite{FlanBorgn}.
It is worth noting that in the present work,  kernel adaptation and sparsity
promotion are dealt with sequentially.
Alternatively (and perhaps more efficiently) \cite{Jokanovic}, 
the function $\sigma(\theta)$ in eq. (\ref{eq:radial}) can be sought 
as part of the solution of the sparsification problem  (\ref{eq:L0}).
% 
%Other, possibily more efficient strategies are possible; 
%e.g., in \cite{Jokanovic} the function $\sigma(\theta)$ in eq. (\ref{eq:radial}) 
%is directly sought as part of the solution of the sparsification problem (\ref{eq:L0}).
%
%%%%%%%%%%%%%%%%%%%%%%%%%%%%%%%%%%%%%%%%%%%%%%%%%%%
\section{Numerical Experiments}
\label{sec:num}
%%%%%%%%%%%%%%%%%%%%%%%%%%%%%%%%%%%%%%%%%%%%%%%%%%%
%
In this section we manage to illustrate by numerical simulations how CS-based 
TF skeleton representations may offer sharper time-frequency resolution 
(which is important for classifying glitches and understanding their origin) 
and substantial de-noising (which is important  for detection purposes). 
A deeper analysis, in particular as regards the statistical properties 
of these representations, will be the subject of future work.\\
In our numerical experiments, the domain $\Omega$  was determined 
as discussed in Section \ref{sec:omega},  
using the public-domain RGK toolbox \cite{RGK}.
The gradient projection algorithm  proposed in \cite{Figueiredo}, 
from  the public-domain GPSR toolbox \cite{GPSR} was used 
to solve the optimization problem (\ref{eq:opt4}), 
using  eq. (\ref{eq:guess})  for the parameter $\lambda$.
All waveforms  
were sampled at $1 \mbox{ KHz}$ in chunks  $512$ samples wide,
unless otherwise stated.
Q-transforms were computed using the toolbox from \cite{Qpack},
and HHT transforms were computed using public domain codes 
from the Mathworks repository.
% 
%%%%%%%%%%%%%%%%%%%%%%%%%%%%%%%%%%%%%%
\subsection{Gabor Molecules}
%%%%%%%%%%%%%%%%%%%%%%%%%%%%%%%%%%%%%%
% 
Simple  Gabor  molecules,  consisting  of  two  SG  waveforms (Gabor atoms)
are perhaps the simplest  {\it multicomponent} transient signals, 
and provide a convenient elementary benchmark 
for a preliminary, controlled comparison among different TF representations.
Here, following \cite{Cohen_sep}, we content ourselves with an operational
definition of a multicomponent transient, as
a signal whose TF skeleton consists of {\it separate} features, i.e., 
blobs or ridges whose TF supports  exhibit  the property that 
their frequency (time)  separation at all times (frequencies) 
is larger than the sum of their local frequency (time) spreads.\\
%
%The Q-transform (with Gaussian windowing function) 
%acts as a matched filter for SG waveforms \cite{Q_Scan}, and thus may be expected 
%to work at its best  with Gabor molecules.
%
The  panels in first (leftmost) column of  Fig.s 2-4  display (from top to bottom) 
the time domain data, frequency spectrum, and fiducial TF skeleton,
which can be computed explicitly from the WV, in this case, 
by subtracting the intermodulation terms.\\
The second column (from left to right) shows  
the three visually best Q-transforms (in a range of $Q$ values between $5$ and $50$).\\
The next (third) column shows the WV distribution, the BJ-smoothed WV,
and the CC skeleton computed using the optimized $\Omega$  domain.\\
Finally, the two panels in the fourth and last column display the HHT 
transform, and its plain (top panel) and IBW-aware (bottom panel, here nicknamed {\it fat}) versions.
We used eq. (17) from \cite{IBW_calc} to estimate the IBW 
and construct the IBW-aware version of the HHT, 
assuming the (instantaneous) amplitude to fall off linearly from
a maximum at the IF, down to zero at the half-bandwidth edges, 
while keeping the total energy (integral of squared instantaneous amplitude 
across the IBW) equal to the squared IA of the original HHT. \\ 
In Fig. 2  the molecule consists of two SG having the same  time width,
but different time centroids, carrier frequencies and peak amplitudes;
in Fig. 3  the two atoms have the same carrier frequency but different time centroids; 
in Fig. 4  they have the same time centroid but different carrier frequencies.\\ 
The CC skeleton computed using the optimized $\Omega$  domain
yields the sharpest TF localization, and goes closest to the actual skeleton
for all cases considered.
The visually best  Q-transform appears less well localized.
The {\it fat} HHT also yields a sharp TF signature for the cases in Figures
2 and 3 - but in the case of atoms with 
coincident time-centroids and {\it close} carrier frequencies (Fig. 4), it fails to 
reproduce the actual skeleton, being fooled by the resulting beat. \\ 
Due to the noted limitation, and for the sake of conciseness, 
in the following Sections we will not include the HHT
among the compared TF representations.  
%
%%%%%%%%%%%%%%%%%%%%%%%%%%%%%%%%%%%%%
\subsection{GW Transients and Glitches}
%%%%%%%%%%%%%%%%%%%%%%%%%%%%%%%%%%%%%%
%
The qualitative conclusions valid for Gabor molecules are essentially confirmed
for realistic waveforms representing both sought GW transients and glitches
of environmental/instrumental origin. \\
Figure 5  refers to a supernova core-collapse GW waveform from \cite{Dimmel}.
The sampling frequency is $16 KHz$ in this case.
Here again the TF skeleton obtained using  the  BJ-optimized $\Omega$  domain
yields the best TF localization, followed by the BJ-smoothed WV.
These two representations display sharp TF signatures of both the "bounce" component 
(the "blob") and the subsequent "ringdown" (the "ridge") in the signal.
The Q-transforms are comparatively less sharp and more blurred, 
and show an evident tradeoff between time and frequency resolution, 
as the Q value is varied. 
The WV is heavily affected by signal-noise intermodulation artifacts.\\
Figures 6a-c display three prototypical glitches from the LIGO 5th Scientific Run, 
from \cite{Saulson} (more glitch skeletons are available in \cite{LIGO_rep}).
Here also the TF skeleton obtained using BJ-optimized $\Omega$  domain
yields the sharpest TF localization. 
The BJ-smoothed WV comes next.
The visually-best Q-transform appears to be close to the TF skeleton obtained 
using the square $\Omega$  domain.\\
The case of  Fig. 6a, showing part of the filtered  file $\# 866273637$
observed in the data channel ({\tt DARM\_ERR}) of  LIGO-Hanford detector (H1) 
is particularly interesting. 
Its optimized TF skeleton  (and BJ-smoothed WV) clearly show
four elementary TF  features ("blobs"), suggesting that this glitch is actually
a {\it multicomponent} waveform (this is also confirmed by atomic decomposition).\\
In the case of the typical filtered power mains glitch sensed by an
environmental magnetometer (LIGO {\tt POWMAG} channel)  shown in Fig. 6b 
all TF representations yield acceptable results, except the TF skeleton obtained using 
the square $\Omega$  domain. This is not surprising, given the highly
anisotropic character of the AF in this case.\\
Figure 7 displays a typical "arch-shaped"  glitch seen  in the GEO-600 data  \cite{Was}.
Such glitches originate from  scattered light  where seismic motion modulates
the scatterers' velocity, and hence represent nonlinear products of environmental noise.
Also in this case,  the  TF  skeleton obtained using  the  BJ-optimized $\Omega$  domain
and the BJ-smoothed WV yield quite sharp TF distributions.
The TF skeleton obtained using the square domain is on a par with
the (visually best) Q-transform in terms of resolution, while looking less noisy
compared to this latter.\\
%
%%%%%%%%%%%%%%%%%%%%%%%%%%%%%%%%%%%%%%
\subsection{TF Consistency Tests}
%%%%%%%%%%%%%%%%%%%%%%%%%%%%%%%%%%%%%%
%
Sparse representations are intrinsically denoising \cite{Donoho}.  
Accordingly we may expect that using  TF skeletons
may boost  the performance of  TF  consistency  tests 
among multiple detector data.\\
This is illustrated in Fig.s 8a to 8h. 
Here we assume for simplicity that the direction of arrival of the GWs 
is fiducially known from different  (e.g.,  EM or neutrino) observations (triggered detection), 
and consider the following simplest TF-consistency test:
\begin{enumerate}
\item{let  $W_m$  the TF representation   (plain or BJ-smoothed WV, or TF estimated
TF skeleton)  built from the data gathered by the $m-$th receiver, $m=1,2,\dots,M$;}
\item{time-shift the TF representations by the appropriate (known) propagation delays;}
\item{compute the "coincidence" TF representation 
\beq
W^{(c)}(t,f)=\left| \prod_{m=1}^M  W_m(t,f) \right|^{1/M};
\eeq
}
\item{define the pixels in the coincidence TF  representation whose  level 
is above the median as "hot".}
\end{enumerate} 
In Fig.s  8a  to 8h  we  restrict ourselves to the simplest  2-detectors case.
In Fig.s  8a and 8b 
the data contain a GW chirp corrupted by glitches in additive white Gaussian noise. 
The SNR of the chirp against the Gaussian background is $15$, 
and is twice that of  the glitches.
Figure 8a displays the WV, the BJ-smoothed WV and the TF skeleton
(using the BJ-optimized $\Omega$ domain).
Figure 8b displays the Q-transforms, for $Q=8$, $20.2$ and $32$.
The panels in the bottom row of Fig. 8a and 8b show the coincidence TF distributions.\\
The intermodulation artifacts of the WV are evident in Fig. 8a, and impair its readability. 
The BJ-smoothed WV, and the  TF skeletons in  Fig. 8a,
and the Q-transforms in Fig. 8b 
clearly show both the GW skeletons  and the offending glitches. 
These latter disappear, as expected, in the coincidence representations.
The  TF skeleton exhibits the sharpest time-frequency localization, 
and also the largest contrast against the background,  
compared to both the  BJ-smoothed WV and the Q-transforms,
suggesting good denoising, further discussed in Sect. \ref{sec:denoising}.\\
In Fig.s  8c and 8d 
the data contain a GW merger corrupted by glitches, in additive white Gaussian noise. 
The SNR of the merger against the Gaussian background is $10$, 
and is twice that of  the glitches.
Figure 8c displays the WV, the BJ-smoothed WV and the TF skeleton
(using the BJ-optimized $\Omega$ domain).
Figure 8d displays the Q-transforms, for $Q=8$, $12.7$ and $20$.
The panels in the bottom row of Fig. 8c and 8d show the pertinent coincidence TF distributions.\\
Inspection of these figures leads to  conclusions
similar to those drawn from Fig.s  8a and 8b.\\
Finally, in Fig.s  8e and 8f the data consist of white Gaussian noise only.
The BJ-smoothed WV and the Q-transforms exhibit some peculiar hot  pixel  patterns. 
The TF skeleton coincidence clearly  features the lowest number of spurious hot  pixels.\\
%
%%%%%%%%%%%%%%%%%%%%%%%%%%%%
\subsection{Coincidence TF Skeletons and Denoising}
\label{sec:denoising}
%%%%%%%%%%%%%%%%%%%%%%%%%%%%
%
The nice properties of the coincidence TF skeleton in terms of denosing
are illustrated in Fig.s 9a and 9b,
showing the level histograms of the "hot" pixels
drawn from the various  coincidence TF representations discussed above.
In each histogram panel the number of hot pixels, their average level,
and their cumulative level sum are displayed.\\
In the WV,  BJ-smoothed WV  and Q-transform cases  
the number of hot pixels is one half the size 
of the corresponding (discrete) TF representations.
The average level and cumulative level sum 
of the hot pixels for the two cases where a signal
is present is only slightly larger (by a factor always $<2$) 
compared to the case of noise only.\\
In the TF skeleton,  as an effect of sparsification,  
the number of hot pixels is smaller by  a factor $\sim\! 10^2$ 
compared to the size of the (discrete) WV. 
In the chirp case (with SNR=15) the energy content of the signal in the TF plane
is spread along a ridge  
(consisting of $\sim\!1.6 \cdot 10^3$ hot pixels, in our example)
while in the merger case (with SNR=10) 
it is concentrated in a blob 
(consisting of $\sim\!3 \cdot 10^2$ hot pixels, in our example).
Compared to the case of noise only,
the average level of the hot pixels  
is  larger by a factor  $\sim 5$ (for the chirp) and $\sim 10$ (for the merger), 
and the cumulative sum of the hot pixel levels 
is  larger by a factor  $\sim\!1.7$ (for the merger)  and $\sim\!4.5$ (for the chirp).\\
The above suggests that TF based skeletonization has inherent denoising
capabilities,  and that  the WV,  the  BJ-smoothed WV and the Q-transform
have comparable performances in TF coincidence tests. 
%
%%%%%%%%%%%%%%%%%%%%%%%%%%%%%%%%%%%%%%%%
\section{Conclusions}
\label{sec:concl}
%%%%%%%%%%%%%%%%%%%%%%%%%%%%%%%%
%
We  reviewed the compressed sensing (CS) paradigm,  
and suggested its possible use 
in TF  GW  data analysis, using illustrative  examples.   
Gravitational wave chirps and bursts as well as instrumental glitches 
share the common feature of being represented by {\it sparse} TF objects,
including  thin $1D-$ ridges, and  localized  $2D-$ juts.
The main goal of the TF skeleton construction algorithm is to produce 
readable TF representations of  complex, sparse,  multi-component waveforms, 
which are free from intermodulation artifacts and  
still offer as much time-frequency resolution as possible.
These are {\it conflicting} requirements,  nonetheless the WV-CS based approach
yields an excellent  tradeoff between them, as illustrated.
We managed to show that TF skeletons  may  yield better insight 
into the {\it fine structure} of  glitches, hopefully improving 
our understanding about their origin, and our skill in classifying them.
We also showed that CS-based sparse TF representations (skeletons) 
act as an effective denoising algorithm, improving
TF-based coincidence tests.\\
We accordingly suggest that sparse CS-based TF skeletons   
may be nice complements  to  existing tools for GW data analysis.
%
%%%%%%%%%%%%%%%%%%%%%%%%%%%%%%%%%%
\section*{Acknowledgements}
%%%%%%%%%%%%%%%%%%%%%%%%%%%%%%%%%%
%
This work has been supported by the Italian Ministry for University and Scientific Research (MIUR) through the grant 20082J7FBN.
%
%++++++++++++++++++++++++++++++++++++++++++

%
\newpage
\newpage


\section*{Supplementary material}

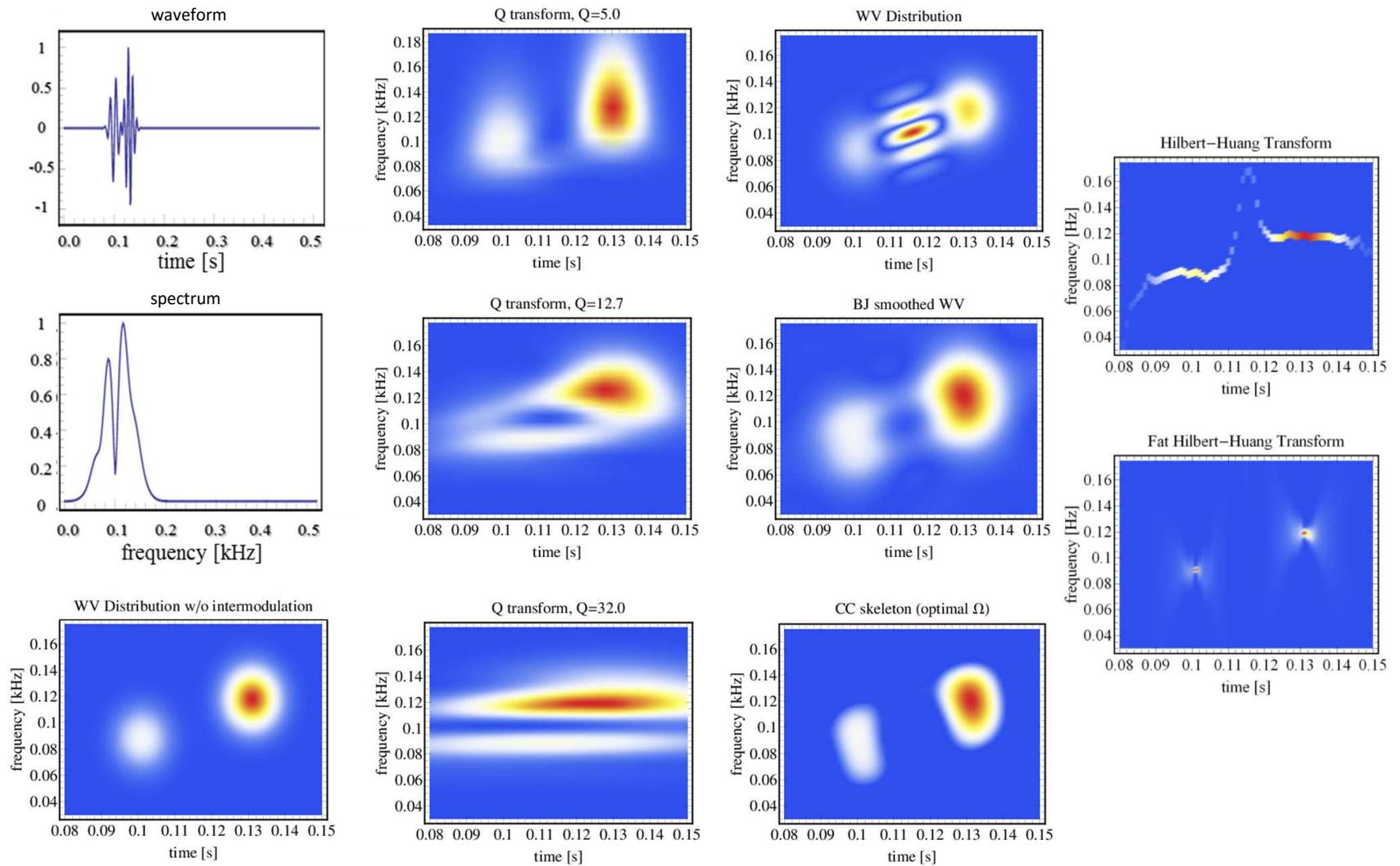

Figure 2 – Gabor molecule consisting of two atoms with different time centroids and carrier frequencies. Comparison among the Q transform, the Wigner-Ville (plain, BJ-smoothed and CC cured), and the Hilbert-Huang transform without and with (fat) IBW information.

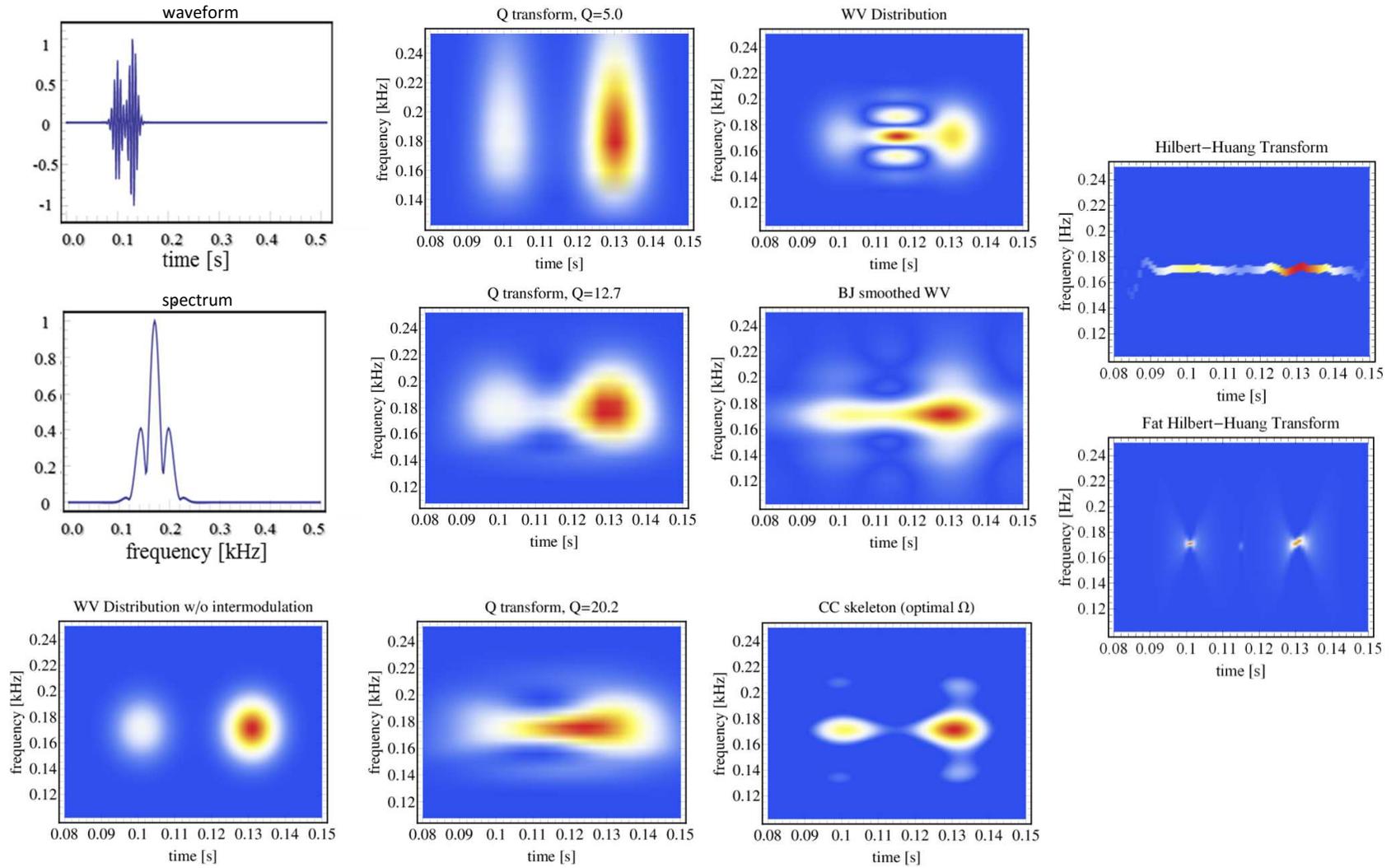

Figure 3 – Gabor molecule consisting of two atoms with different time centroids and equal carrier frequencies. Comparison among the Q transform, the Wigner-Ville (plain, BJ-smooth- ed and CC cured), and the Hilbert-Huang transform without and with (fat) IBW information.

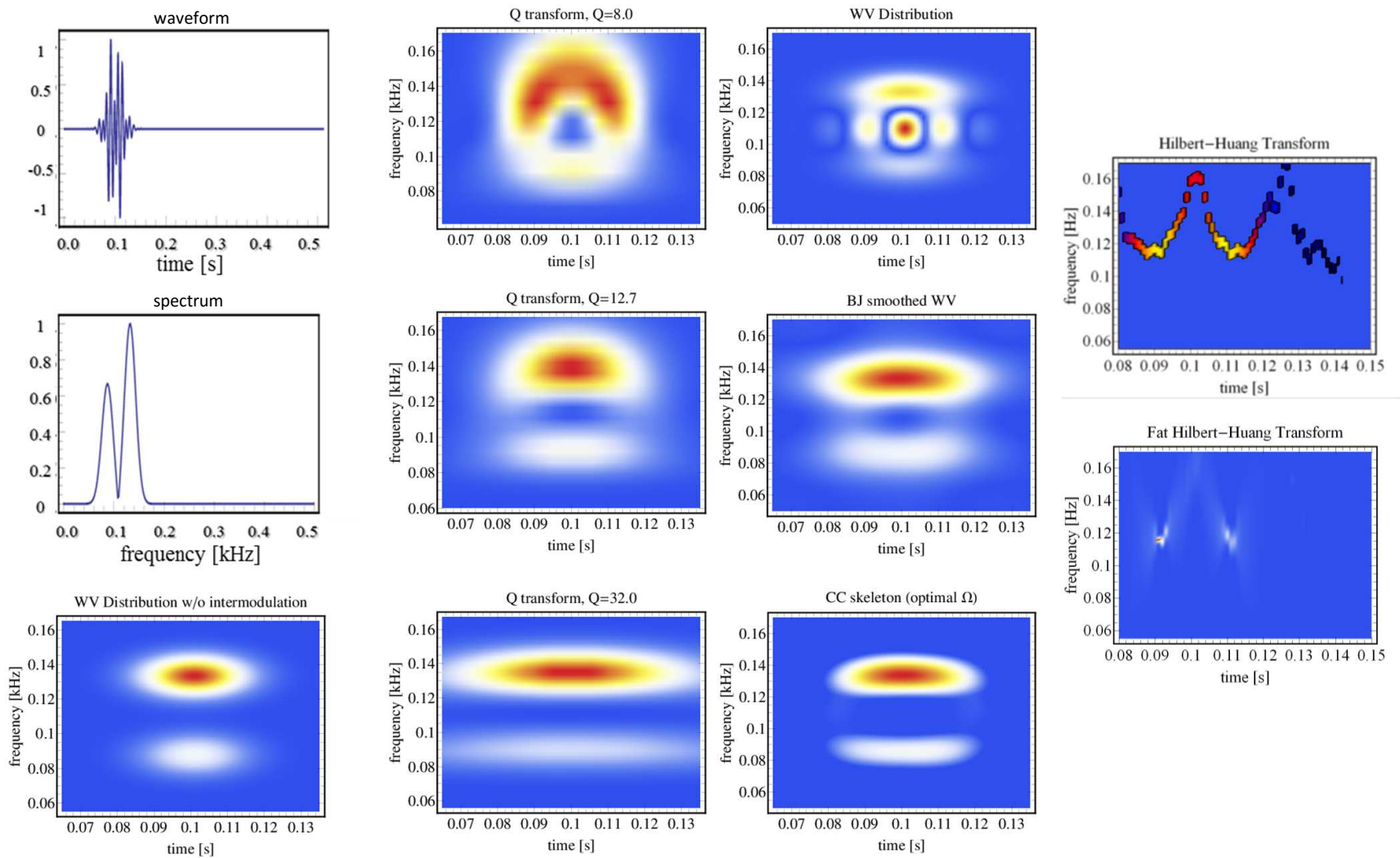

Figure 4 – Gabor molecule consisting of two atoms with equal time centroids and different carrier frequencies. Comparison among the Q transform, the Wigner-Ville (plain, BJ-smooth-ed and CC cured), and the Hilbert-Huang transform without and with (fat) IBW information.

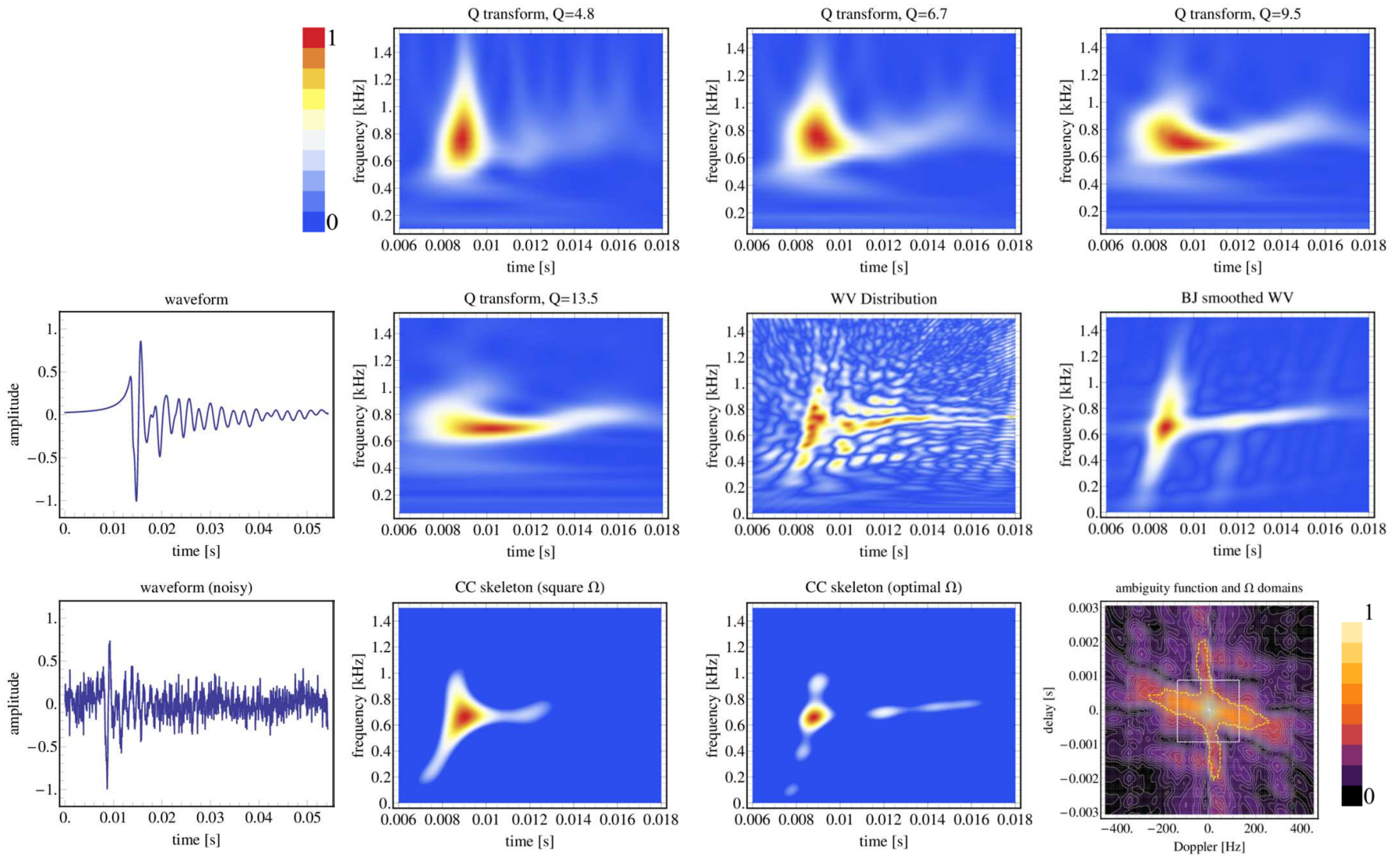

Figure 5 - Supernova core-collapse GW waveform from [31]. The panels display the time waveform (noise-free and with added white Gaussian noise), the Q-transforms for Q=4.8, 6.7, 9.5, and 13.5, the WV and the BJ-smoothed WV, the CC skeletons obtained using the square and BJ-optimized Ω domain, and the pertinent ambiguity function, together with the square and BJ-optimized Ω domain boundaries.

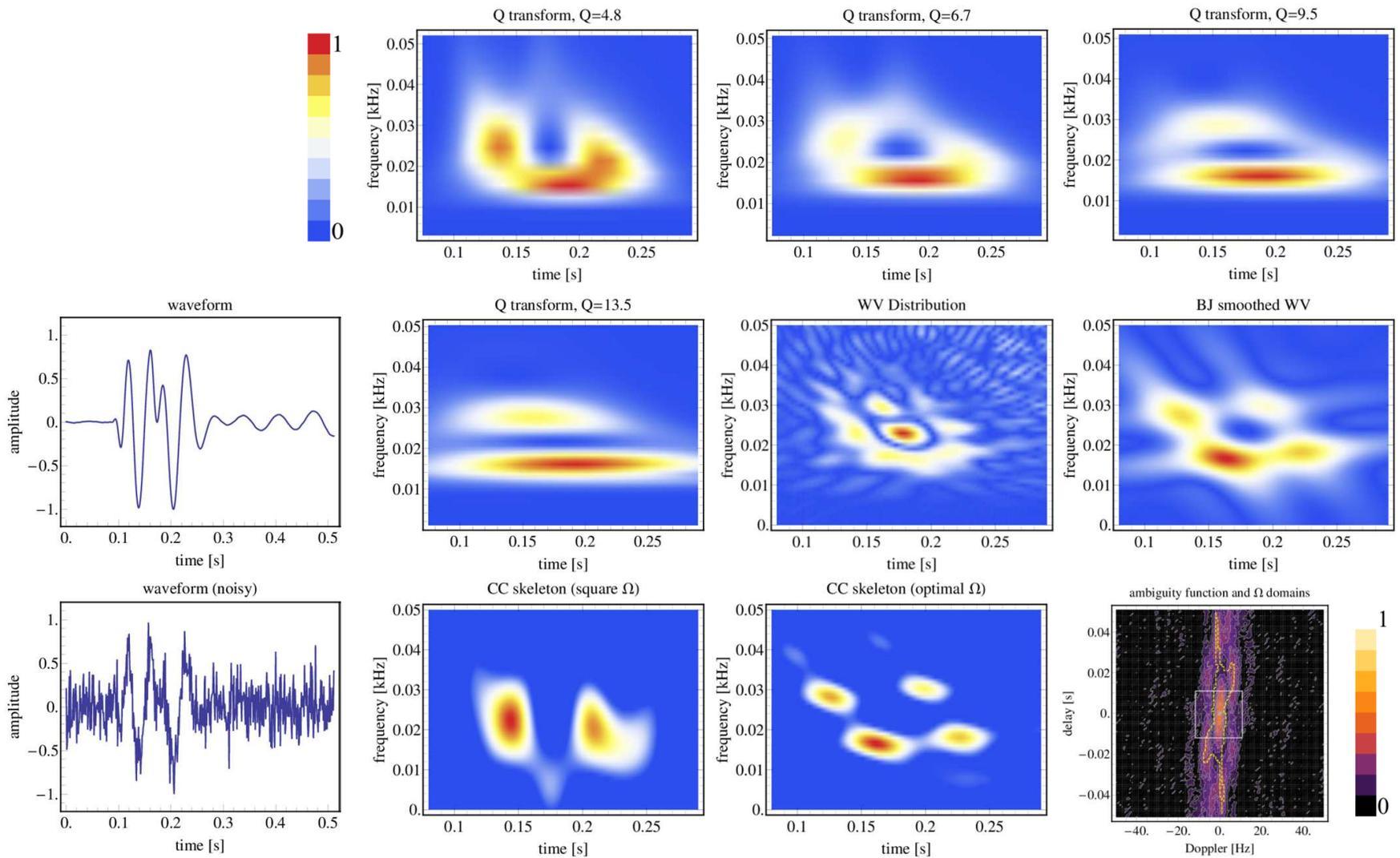

Figure 6a - A LIGO glitch from the 5th Science Run (part of H1 DARM_ERR_866273637 from [51]). The panels display the time waveform (noise-free and with added white Gaussian noise), the Q-transforms for Q=4.8, 6.7, 9.5, and 13.5, the WV and the BJ-smoothed WV, the CC skeletons obtained using the square and BJ-optimized Ω domain, and the pertinent ambiguity function, together with the square and BJ-optimized Ω domain boundaries.

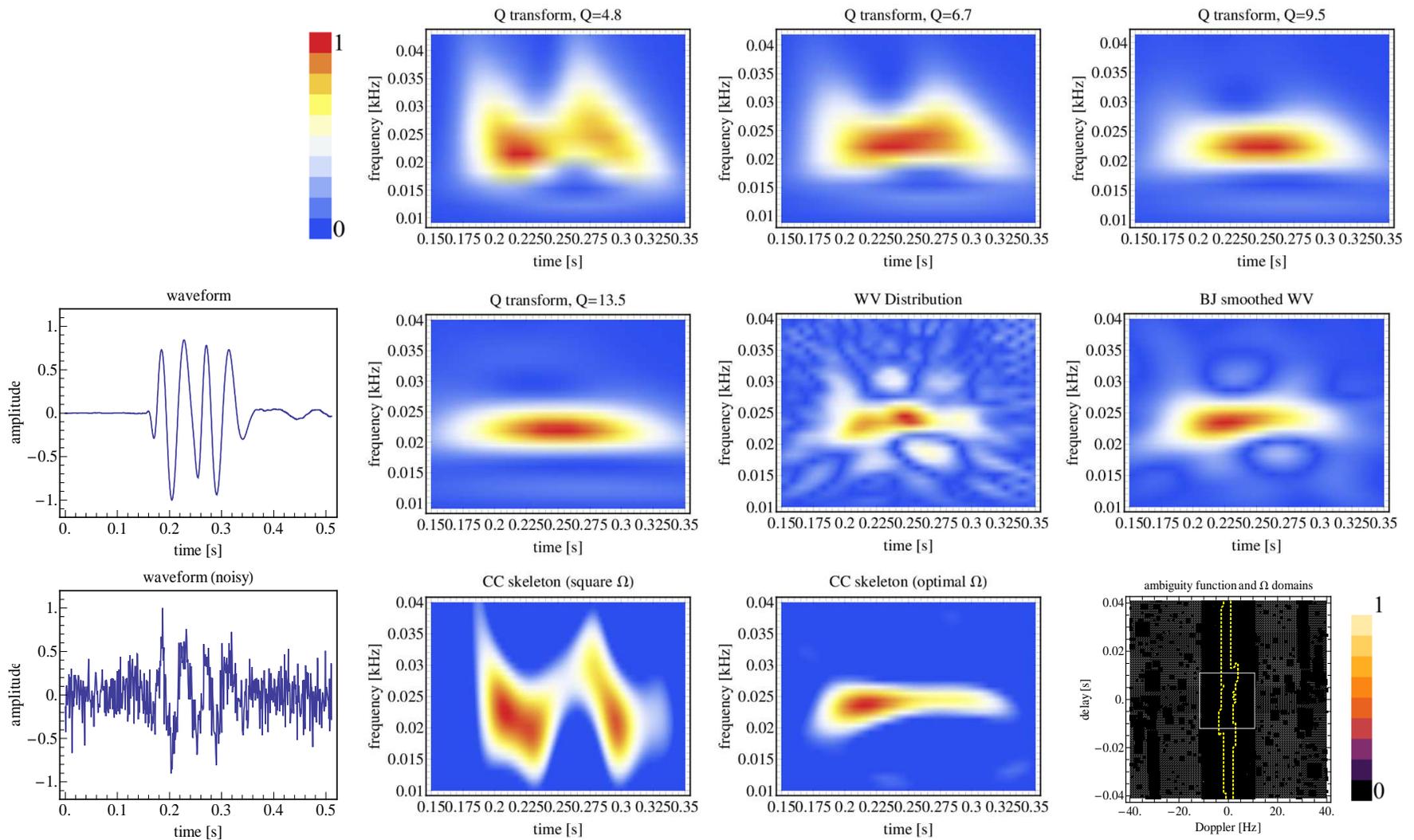

Figure 6b - A LIGO (`power_mag`) glitch from the 5th Science Run from [51]. The panels display the time wave-form (noise-free and with added white Gaussian noise), the Q-transforms for Q=4.8, 6.7, 9.5, and 13.5, the WV and the BJ-smoothed WV, the CC skeletons obtained using the square and BJ-optimized $\Omega$ domain, and the pertinent ambiguity functions, together with the square and BJ-optimized $\Omega$ domain boundaries.

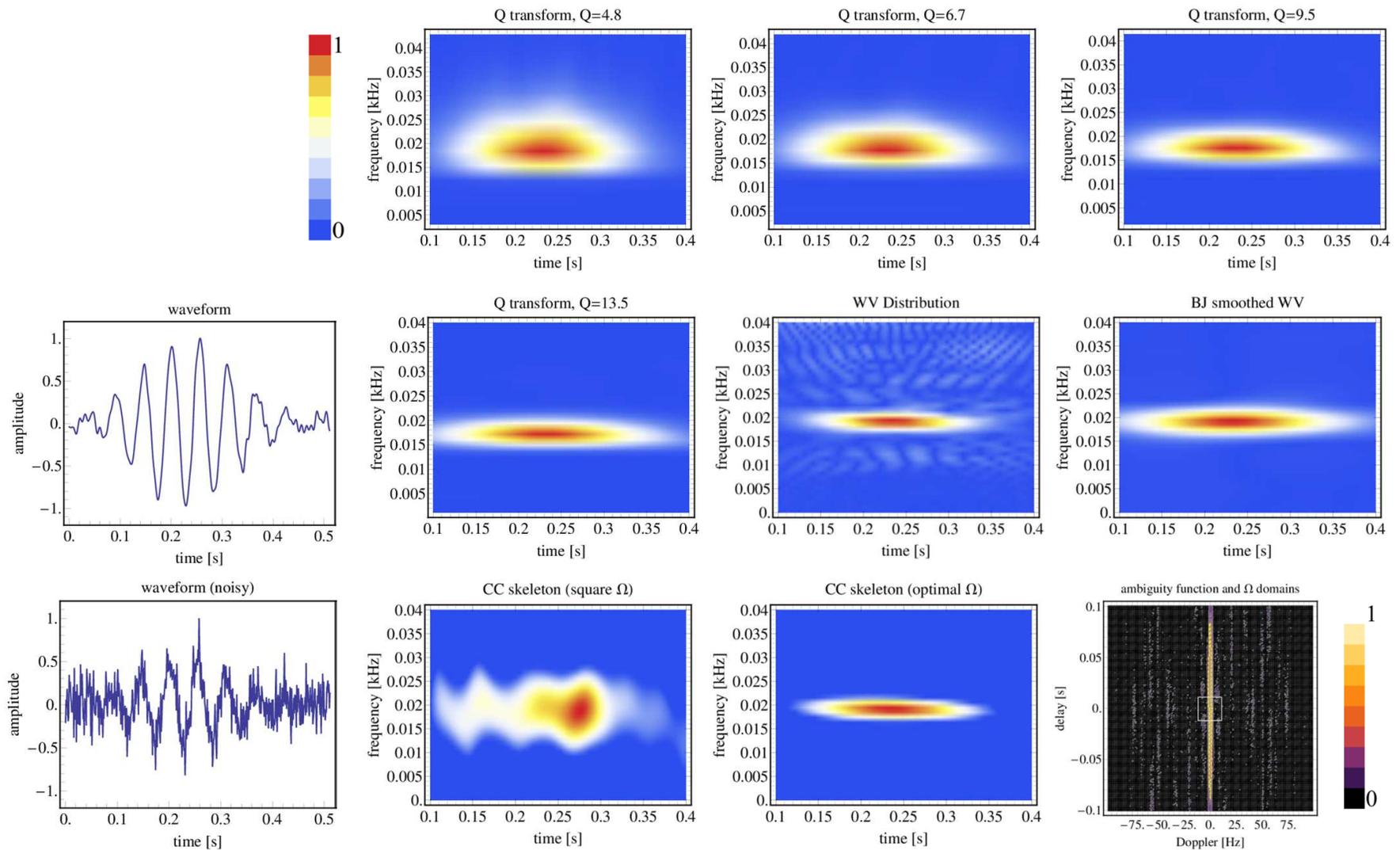

Figure 6c - A LIGO (`power_mag`) glitch from the 5th Science Run from [51]. The panels display the time wave-form (noise-free and with added white Gaussian noise), the Q-transforms for Q=4.8, 6.7, 9.5, and 13.5, the WV and the BJ-smoothed WV, the CC skeletons obtained using the square and BJ-optimized Ω domain, and the pertinent ambiguity function, together with the square and BJ-optimized Ω domain boundaries.

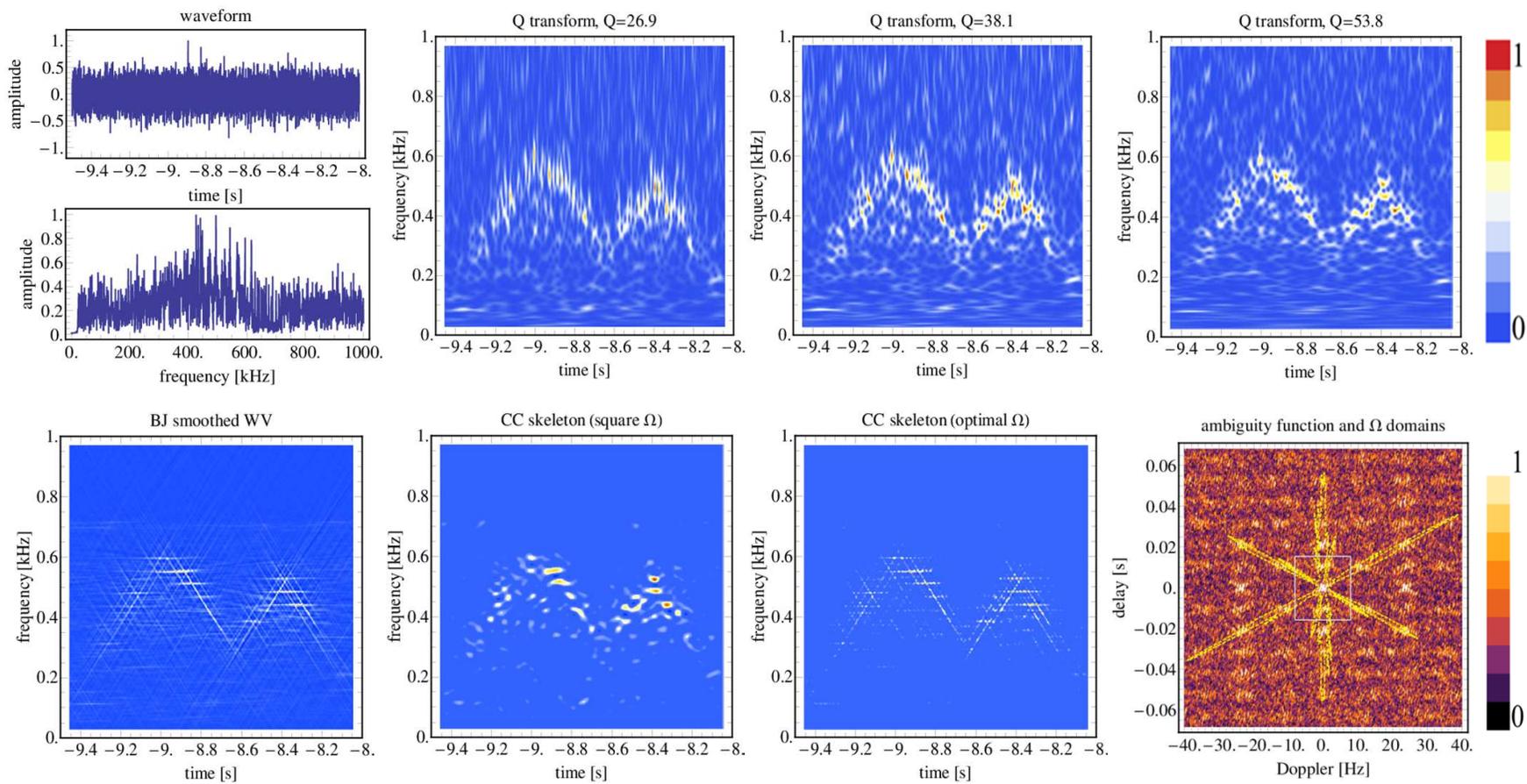

Figure 7 - A GEO-600 arch-glitch [51]. The panels display the (noisy) waveform and its frequency spectrum, the Q-transforms for Q=26.9, 38.1 and 53.8, the BJ-smoothed WV, the CC skeletons obtained using the square and BJ-optimized Ω domain, and the pertinent ambiguity function, together with the square and BJ-optimized Ω domain boundaries.

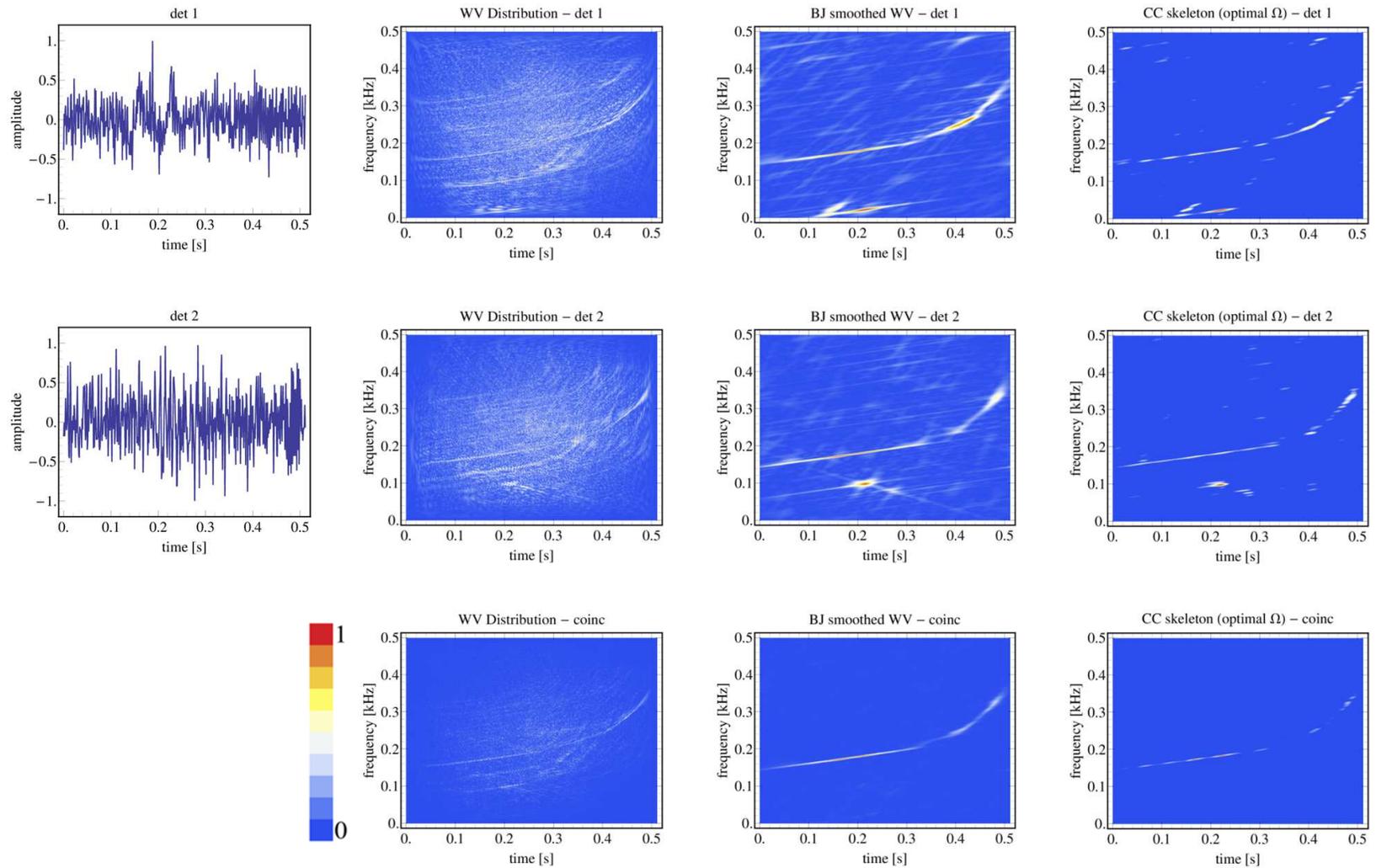

Figure 8a - TF Consistency tests (triggered search, incoherent detection) between two detectors, eq. (24). The panels display the waveforms and the WV, BJ-smoothed WV, and CC skeleton with BJ-optimized Ω domain. The data in each detector include a Newtonian GW chirp with SNR=15 against the additive white Gaussian background, and a random SG-glitch, with SNR=7.5.

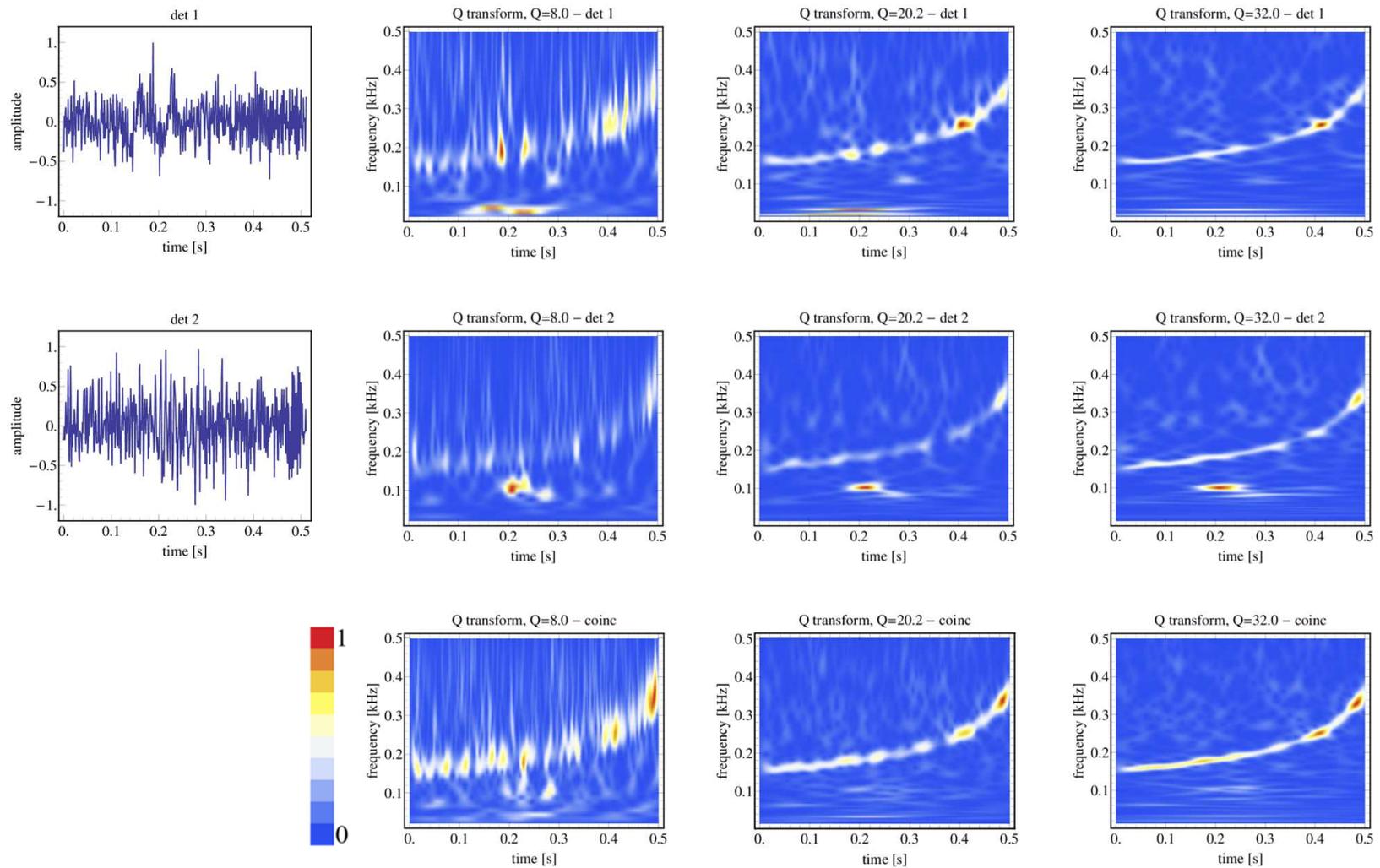

Figure 8b - TF Consistency tests (triggered search, incoherent detection) between two detectors, eq. (24). The panels display the waveforms and the Q-transforms with Q=8, 20.2 and 32. The data in each detector are the same as in Fig.5a, and include a Newtonian GW chirp with SNR=15 against the additive white Gaussian background, and a random SG-glitch, with SNR=7.5.

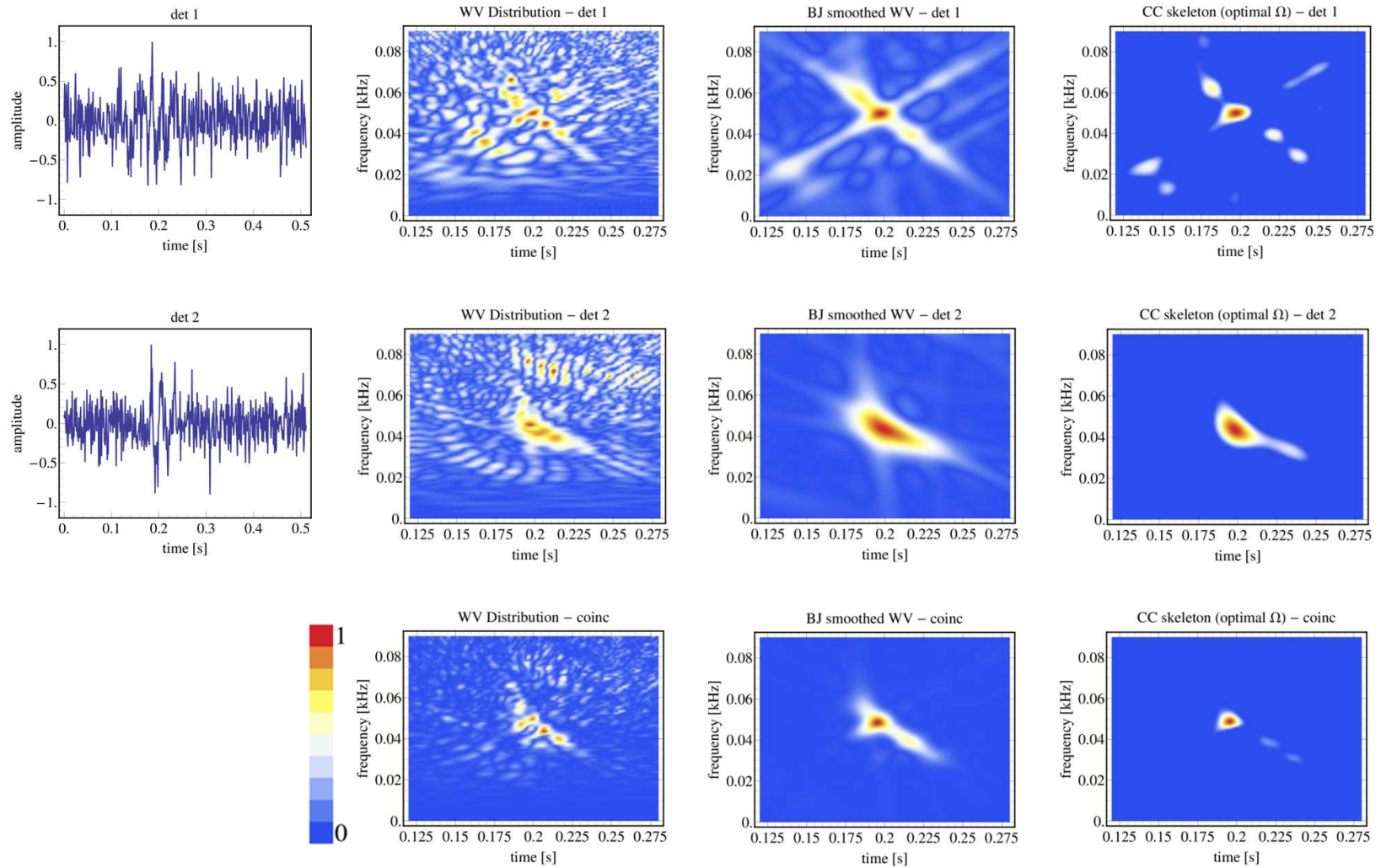

Figure 8c - TF Consistency tests (triggered search, incoherent detection) between two detectors, eq. (24). The panels display the Waveforms and the WV, BJ-smoothed WV, and CC skeleton with BJ-optimized Ω domain. The data in each detector include a GW merger from [30] with SNR=10 against the additive white Gaussian background, and a random SG-glitch, with SNR=5.

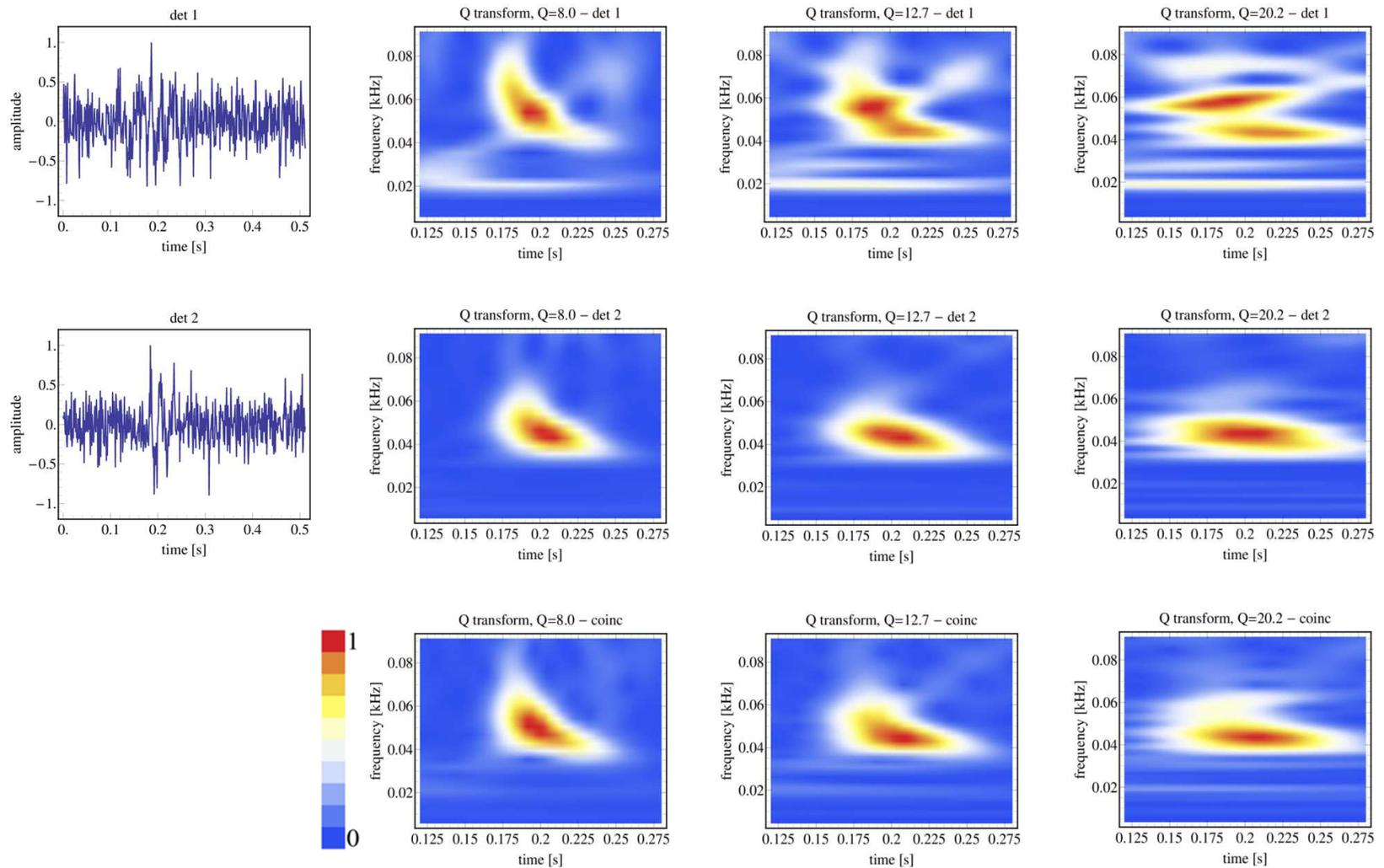

Figure 8d - TF Consistency tests (triggered search, incoherent detection) between two detectors, eq. (24). The panels display the waveforms and the Q-transforms with Q=8, 20.2 and 32. The data in each detector are the same as in Fig.5c, and include a GW merger from [30] with SNR=10 against the additive white Gaussian background, and a random SG-glitch, with SNR=5.

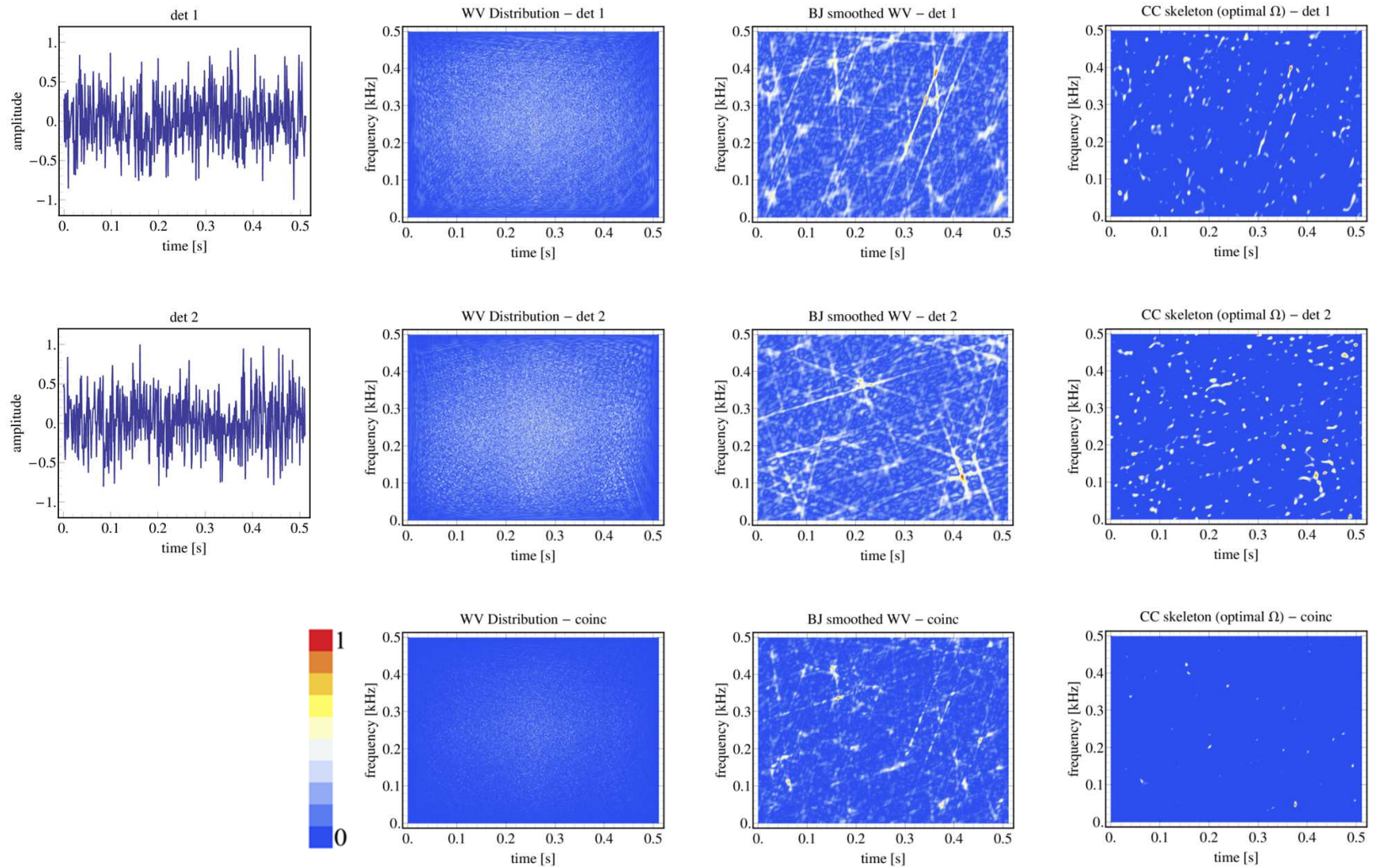

Figure 8e - TF Consistency test (triggered search, incoherent detection) between two detectors, eq. (24). The panels display the waveforms and the WV, BJ-smoothed WV, and CC skeleton with BJ-optimized Ω domain. The data in each detector consist of independent realizations of white Gaussian noise with the same power spectral density as the background noise in Fig.s 5a-5d.

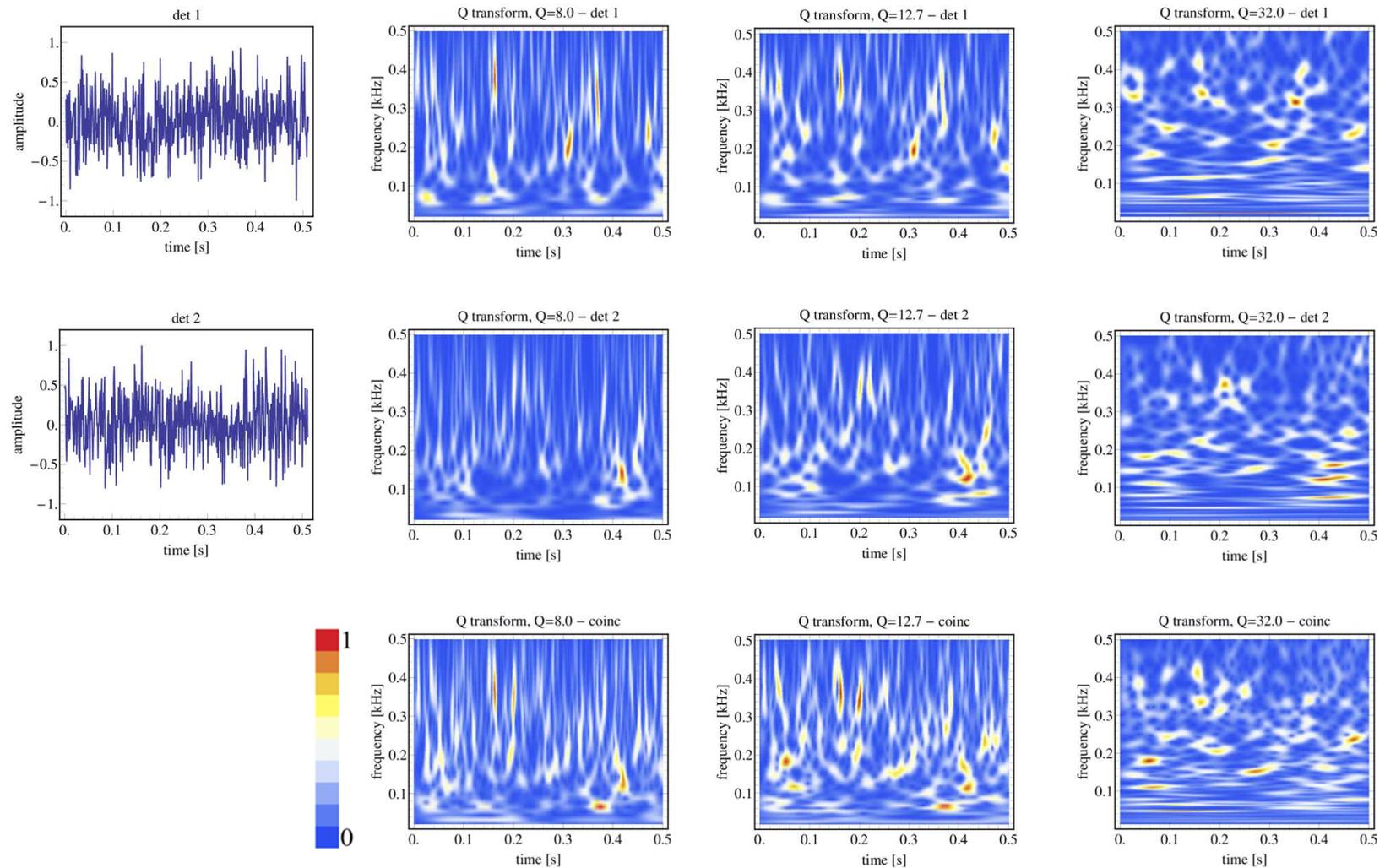

Figure 8f - TF Consistency tests (triggered search, incoherent detection) between two detectors, eq. (24). The panels display the waveforms and the Q-transforms with Q=8, 12.7 and 32. The data in each detector consist of independent realizations of white Gaussian noise with the same power spectral density as the background noise in Fig.s 5a-5d.

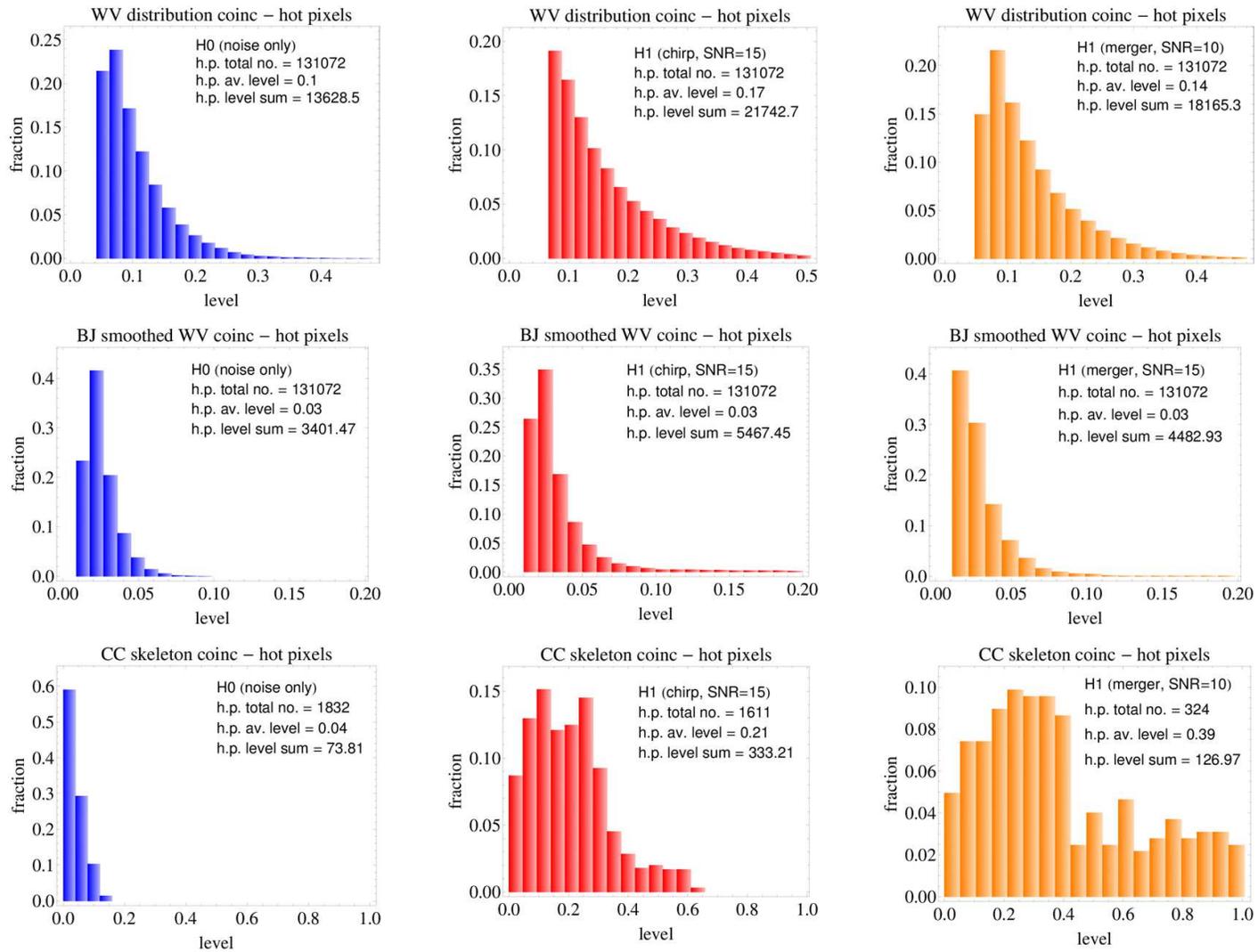

Figure 9a - TF Consistency tests (triggered search, incoherent detection) between two detectors. WV, BJ-smoothed WV, and CC skeleton with BJ-optimized domain. Level histograms of the hot pixels in the coincidence distributions.

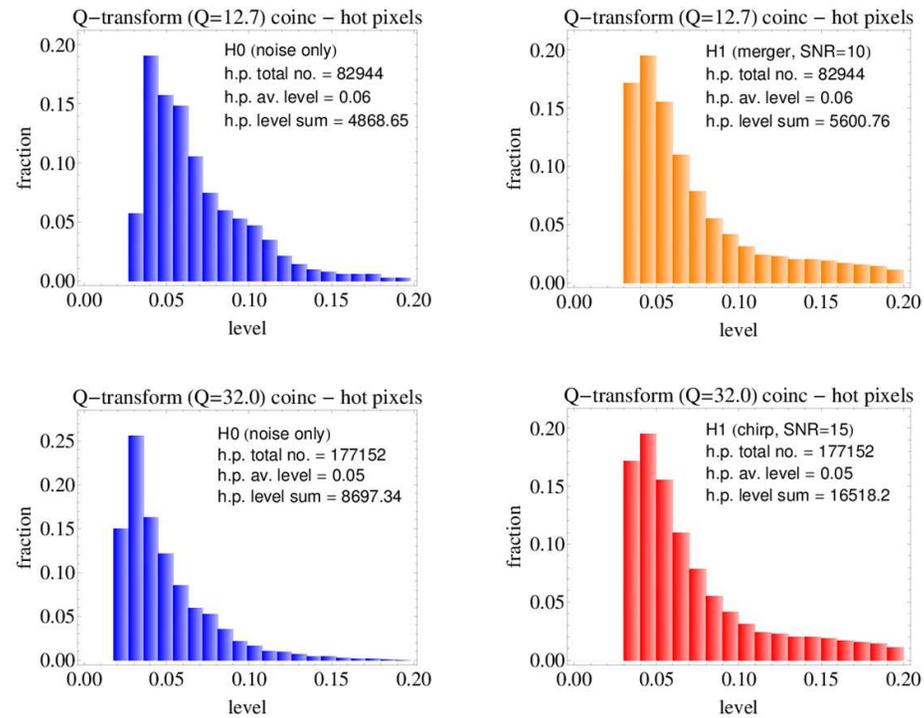

Figure 9b - TF Consistency tests (triggered search, incoherent detection) between two detectors. Q transforms with Q=12.7 (GW merger, SNR=10) and Q=32 (GW chirp, SNR=15). Level histograms of the hot pixels in the coincidence distributions.




\begin{thebibliography}{99}
\bibitem{footnote3}{The equality between the time (frequency) marginal distribution of the WV
and the power (spectral density) of its argument is a special case of the above property.}
\bibitem{footnote2}{The isotropic $\Omega$ domain in the discrete TF plane
is a rectangle whose aspect ratio depends on the chosen  time-window width 
and sampling frequency.}
\end{thebibliography}
\end{document}